\documentclass[fleqn,usenatbib]{mnras}

\usepackage{newtxtext,newtxmath}
\usepackage[T1]{fontenc}

\DeclareRobustCommand{\VAN}[3]{#2}
\let\VANthebibliography\thebibliography
\def\thebibliography{\DeclareRobustCommand{\VAN}[3]{##3}\VANthebibliography}

\usepackage{graphicx}	% Including figure files
\usepackage{subcaption}
\usepackage{amsmath}	% Advanced maths commands
\usepackage{bm}
\usepackage{color}

\newcommand{\diffop}{\mathrm{d}}
\newcommand{\eps}{\epsilon}
\newcommand{\mmat}[1]{{\mathbf{#1}}}
\newcommand{\mP}{\mathcal{P}}
\newcommand{\mD}{\mathcal{D}}

\newcommand{\mL}{\mathcal{L}}
\newcommand{\mX}{\mvec{X}}

\newcommand{\mvec}[1]{{\bm{#1}}}
\newcommand{\karmma}{\texttt{KaRMMa}}
\newcommand{\pp}{$P$-$P$~}

\title[WL map-based inference]{Map-based cosmology inference with lognormal cosmic shear maps}

\author[S. S. Boruah, E. Rozo and P. Fiedorowicz]{Supranta S. Boruah$^{1}$\thanks{Contact e-mail: \href{mailto:ssarmabo@email.arizona.edu}{ssarmabo@email.arizona.edu}}, Eduardo Rozo$^{2}$\thanks{Contact e-mail: \href{mailto:erozo@email.arizona.edu}{erozo@email.arizona.edu}}
Pier Fiedorowicz$^{2}$\thanks{Contact e-mail: \href{mailto:pierfied@email.arizona.edu}{pierfied@email.arizona.edu}}
\\
$^{1}$ Department of Astronomy and Steward Observatory, University of Arizona, 933 N Cherry Ave, Tucson, AZ 85719, USA \\
$^{2}$Department of Physics, University of Arizona, 1118 E. Fourth Street, Tucson, AZ, 85721, USA
}

\date{Accepted XXX. Received YYY; in original form ZZZ}

\pubyear{2022}
\begin{document}
\label{firstpage}
\pagerange{\pageref{firstpage}--\pageref{lastpage}}
\maketitle

\begin{abstract}
Most cosmic shear analyses to date have relied on summary statistics (e.g. $\xi_+$ and $\xi_-$).  These types of analyses are necessarily sub-optimal, as the use of summary statistics is lossy. In this paper, we forward-model the convergence field of the Universe as a lognormal random field conditioned on the observed shear data.  This new map-based inference framework enables us to recover the joint posterior of the cosmological parameters and the convergence field of the Universe. Our analysis properly accounts for the covariance in the mass maps across tomographic bins, which significantly improves the fidelity of the maps relative to single-bin reconstructions.  We verify that applying our inference pipeline to Gaussian random fields recovers posteriors that are in excellent agreement with their analytical counterparts. At the resolution of our maps --- and to the extent that the convergence field can be described by the lognormal model --- our map posteriors allow us to reconstruct \it all \rm summary statistics (including non-Gaussian statistics). We forecast that a map-based inference analysis of LSST-Y10 data can improve cosmological constraints in the $\sigma_8$--$\Omega_{\rm m}$ plane by $\approx 30\%$ relative to the currently standard cosmic shear analysis. This improvement happens almost entirely along the $S_8=\sigma_8\Omega_{\rm m}^{1/2}$ directions, meaning map-based inference fails to significantly improve constraints on $S_8$.
\end{abstract}


%%%%%%%%%%%%%%%%%%%%%%%%%%%%%%%%%%%%%%%%%%%%%%%%%%

%%%%%%%%%%%%%%%%% BODY OF PAPER %%%%%%%%%%%%%%%%%%
\section{Introduction}

Weak lensing is the distortion of galaxy shapes by the large-scale structure in their foreground. Due to its sensitivity to large scale structures at different redshifts, weak lensing observations can be used to constrain the cosmological parameters governing structure growth. It is therefore one of the key probes used to understand the nature of dark matter and dark energy. The current stage-III surveys \citep{Troxel2018, Hamana2020, Heymans2021} and the upcoming stage-IV surveys such as the Rubin Observatory\footnote{\url{https://www.lsst.org/}}, Roman Observatory\footnote{\url{https://roman.gsfc.nasa.gov/}} and Euclid\footnote{\url{https://www.euclid-ec.org/}} will probe the Universe with weak lensing with exquisite precision.

Traditional cosmological analysis of galaxy shear data typically compress the data into two-point statistics, i.e. either the correlation function \citep{Troxel2018} or the power spectrum \citep{Hikage2019}. However, two-point statistics are sub-optimal unless the underlying density field is Gaussian.  Because non-linear gravitational growth introduces non-Gaussian features in the density field, extracting this non-Gaussian information can provide tighter constraints on the cosmological parameters. For example, using non-Gaussian statistics can substantially tighten the constraints on neutrino masses \citep{Liu2019, Marques2019, Ajani2020, Boyle2021} and the dark energy equation of state \citep{Martinet2021}. As we move to Stage-IV cosmic shear surveys, the reduced shape noise in these future data sets will further enhance the relative utility of these non-Gaussian features. Current methods for capturing non-Gaussian information in weak lensing data include: the weak lensing one point PDF \citep{Liu2019, Thiele2020, Boyle2021}; 3 point statistics \citep{Takada2004, Semboloni2011, Halder2021}; Minkowski functionals \citep{Kratochvil2012, Petri2013}; moments of mass maps \citep{Gatti2020, Gatti2021}; peak counts \citep{Liu2015, HarnoisDeraps2021, Zurcher2022}; and wavelet transforms, \citep{Cheng2020, Cheng2021, Ajani2021} among others. Recently, machine learning based methods have also been proposed as tools for extracting the non-Gaussian information content in weak lensing data \citep{Ribli2019, Fluri2019, Matilla2020, Jeffrey2021, Fluri2022}. 

We propose an alternate method to extract the non-Gaussian cosmological information directly from the mass maps at the field level. Our method relies on a Bayesian forward-modelling framework. This type of field level forward-modelled analysis methods have recently gained increasing prominence in the literature \citep[e.g. ][]{Jasche2013, Millea2020b, Boruah2021}. In \citet{Leclercq2021}, it was shown using a toy model that a field level analysis is both more precise and accurate than a two-point function analysis. Such field level analysis methods have also been proposed for weak lensing. For example, \citet{Alsing2016, Alsing2017} proposed a Bayesian forward modelling framework to infer cosmological parameters from shear maps assuming a Gaussian prior on the 2D convergence field. \citet{Bohm2017} used a lognormal prior on the 3D density field to forward-model the convergence field. Recently, in \citet{Porqueres2021, Porqueres2022}, the full 3D density field was forward modelled and reconstructed using the {\sc borg} algorithm \citep{Jasche2013, Jasche2019}. 

In this paper, we extend the framework of the \karmma\ algorithm we developed in \citet{Fiedorowicz2022} to enable map-based inference of cosmological parameters.  The \karmma\ algorithm samples the real-space values of the convergence field subject to a log-normal prior and conditioned on cosmic shear data.  We have already shown \karmma\ maps are nearly unbiased and accurately reproduce the 1-point and 2-point statistics of the convergence field, as well as the peak and void statistics of the same.  These successes mark a dramatic improvement relative to other techniques that have been successfully applied to data.  However, the original \karmma\ algorithm suffered from three critical drawbacks: 1) the computational and memory requirements associated with storing the covariance matrix of the density field limited the original \karmma\ maps to relatively coarse resolution; 2) mass maps were reconstructed a single tomographic bin at a time, and therefore ignored the correlation between them; and 3) the cosmological parameters governing the log-normal prior were held fixed during sampling.  In this paper, we demonstrate how to simultaneously address all three drawbacks to enable joint sampling of correlated tomographic mass maps and cosmological parameters.  At the resolution of our maps and to the extent that convergence maps can be described by lognormal fields, the method presented here is optimal for inferring cosmological parameters, in the sense that all summary statistics are included in the map.

This paper is structured as following: we begin with the essential weak lensing background in Section \ref{sec:background}. In Section \ref{sec:methods} we describe our forward modelled map-based inference method. In section \ref{sec:performance_test} we validate our code by comparing the results of our method with analytic results. In Section \ref{sec:results}, we describe the results of using our code on simulated data. We then conclude in Section \ref{sec:conclusion}. Throughout this paper, we use the flat sky approximation.

\section{Theory background}\label{sec:background}

\subsection{Weak lensing Theory}\label{ssec:wl_theory}

Galaxy shapes are noisy observations of the underlying cosmic shear field, $\gamma$. The shear field is in turn related to the convergence, $\kappa$. The convergence along a given direction in the sky is sensitive to the matter field in the foreground of a source galaxy. The convergence field in the $i$-th tomographic bin can be written as
\begin{equation}
    \kappa^{(i)}(\mvec{\theta}) = \int_{0}^{\chi_{\text{H}}} d \chi\ g^{(i)}(\chi) \delta(\chi \mvec{\theta}, \chi),
\end{equation}
where, $g^{(i)}(\chi)$ is the lensing efficiency for the $i$-th tomographic bin, 
\begin{equation}
    g^{(i)}(\chi) = \frac{3 H^2_0 \Omega_m}{2 c^2}\frac{\chi}{a(\chi)} \int_{\chi}^{\chi_{\text{H}}} d \chi^{\prime} \frac{\chi^{\prime} - \chi}{\chi^{\prime}} p^{(i)}(\chi^{\prime}).
\end{equation}
Here, $p^{(i)}(\chi)$ is the source redshift probability distribution of the galaxies in the $i$-th tomographic bin, $\chi$ denotes the comoving distance and $\chi_{\text{H}}$ is the comoving horizon distance. The above expression assumes a spatially flat Universe. The shear field is related to the convergence field through the Kaiser-Squires transformation. In order to avoid confusion with Kaiser-Squires inversion \citep{Kaiser1993}, we call the convergence-to-shear relation, the `forward' Kaiser-Squires relation. The forward Kaiser-Squires relation relates the Fourier transform of the convergence field to the Fourier transform of the two components of the shear fields via
\begin{align}\label{eqn:fwd_KS}
    \tilde{\gamma}^{1(i)}_{\mvec{l}} &= \frac{l^2_x - l^2_y}{l^2} \tilde{\kappa}^{(i)}_{\mvec{l}},\\
    \tilde{\gamma}^{2(i)}_{\mvec{l}} &= \frac{2 l_x l_y}{l^2}\tilde{\kappa}^{(i)}_{\mvec{l}}.
\end{align}
In our notation, we denote our Fourier modes with $\tilde{[.]}$.  These are related to the real space field via
\begin{equation}\label{eqn:kappa_power_spectrum}
    \tilde{\kappa}^{(i)}_{\mvec{l}} = \frac{1}{\Omega_s}\int_{\Omega_s} d^2 \mvec{\theta}\ e^{-i \mvec{l}\cdot \mvec{\theta}} \kappa^{(i)}(\mvec{\theta}),
\end{equation}
where, $\Omega_s$ denotes the survey area on the sky. Due to homogeneity and isotropy, the covariance of the Fourier modes is diagonal, i.e,
\begin{equation}\label{eqn:kappa_cov_gaussian}
    \langle \tilde{\kappa}^{(i)}_{\mvec{l}} \tilde{\kappa}^{(j)}_{\mvec{l}^{\prime}}\rangle = \Omega_s C^{(ij)}_{l} \delta^{(K)}_{\mvec{l}, -\mvec{l}^{\prime}},
\end{equation}
where, $C^{(ij)}_{l}$ is the (cross-) correlation for the Fourier modes of convergence between the $i$-th and the $j$-th tomographic bin and $\delta^{(K)}$ is the Kronecker delta symbol.

\subsection{The Lognormal Model}

In cosmological applications, non-Gaussian fields are commonly modelled as lognormal fields. This includes the modelling of the 3D density field \citep{Coles1991, Jasche2010, Bohm2017} as well as the 2D convergence field \citep{Taruya2002, Clerkin2017}. %However, w
While the lognormal model does not approximate the 3D density field well \citep{Klypin2018} it does accurately describe the 2D convergence field \citep{Clerkin2017,Xavier2016}. 

In the lognormal model the convergence field of the $i$-th redshift bin, $\kappa^i$, is an exponential transformation of a Gaussian field, $y_i$. The two variables are related via
\begin{equation}
    \kappa^i(\mvec{\theta}) = e^{y_i(\mvec{\theta})} - \lambda_i,
\end{equation}
where $\lambda_i$ is referred to as the shift parameter of the lognormal distribution. The shift parameter depends on the scale at which the field is smoothed/pixelized. Once the shift parameter is determined, the correlation function of $y$ is related to the correlation function of $\kappa$ via
\begin{equation}
    \xi^{ij}_{y}(\theta) = \log\bigg[\frac{\xi^{ij}_{\kappa}(\theta)}{\lambda_i \lambda_j} + 1
    \bigg].
\end{equation}
In Fourier space, the power spectrum of the variable $y$ is 
\begin{equation}
    C^{ij}_y(\ell) = 2\pi \int_{0}^{\pi} \diffop \theta\ \sin \theta P_{\ell}(\cos \theta)\xi^{ij}_{y}(\theta),
    \label{eq:PS}
\end{equation}
where $P_{\ell}$ is the Legendre polynomial of order $\ell$. The covariance of the variable $y$ is diagonal and given by
\begin{equation}\label{eqn:y_cov}
    \langle \tilde{y}^{(j)}_{\mvec{l}} \tilde{y}^{(k)}_{\mvec{l}^{\prime}}\rangle = \Omega_s C^{(jk)}_{y,l} \delta^{(K)}_{\mvec{l}, -\mvec{l}^{\prime}}.
\end{equation}
Thus, the non-Gaussian convergence field can be modelled using a deterministic transformation of the Gaussian $y$ variables. One of the important features of the lognormal model is that it allows us to simultaneously constrain the convergence fields in multiple tomographic bins while accounting for the covariance between the bins, as is evident from equation~\ref{eq:PS}.
\section{Map-based cosmology inference and forward-modelled map reconstruction}\label{sec:methods}

The objective of our analysis is to simultaneously sample the convergence maps $\mvec{\kappa}$ and cosmological parameters $\Theta$ given the weak lensing data $\mD$.  More precisely, we want to sample from the posterior, $\mP(\mvec{\kappa}, \Theta| \mD)$. We do so using a block sampling scheme: we first sample from $\mP(\mvec{\kappa}|\Theta, \mD)$, i.e. we sample the convergence maps given a set of cosmological parameters $\Theta$. In the second step, the updated mass maps are held fixed while the cosmological parameters are sampled from the conditional posterior $\mP(\Theta|\mvec{\kappa}, \mD)$.

\subsection{Map sampling}

\subsubsection{Reparameterization of $\kappa$}\label{sssec:kappa_reparam}
\begin{figure*}
    \centering
    \includegraphics[width=\linewidth]{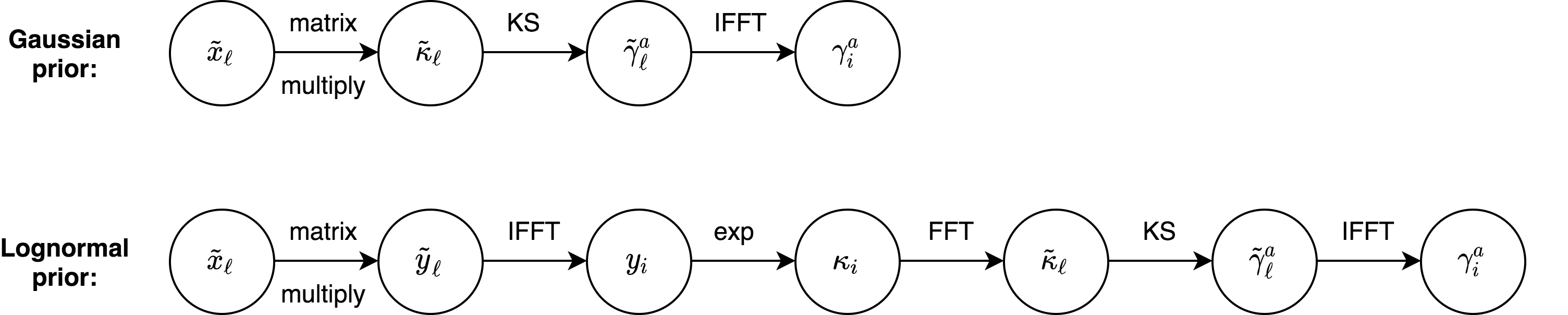}
    \caption{An illustration of the steps involved in the transformation of the $x_{\mvec{l}}$ variable to pixelized shear. Here, `KS' refers to the forward Kaiser-Squires transformation, `FFT' is the Fourier transform, `IFFT' is the inverse Fourier transform, `exp' is exponentiation. Note that the reparameterization technique described in section \ref{sssec:kappa_reparam} works on different variables for the cases of the Gaussian prior and the lognormal prior. While sampling using our HMC code, the gradient of the likelihood is calculated using automatic differentiation which is propagated back through each of the involved steps.}
    \label{fig:fwd_model}
\end{figure*}

We find that naive block sampling of the convergence fields is highly inefficient due to strong degeneracies between the $\mvec{\kappa}$ field and the cosmological parameters.  Fortunately, this is easily overcome through a reparameterization of the $\mvec{\kappa}$ field.  Specifically, we perform an eigenvalue decomposition of the power spectrum  of the convergence field, $C_{\mvec{l}}$, as,
\begin{equation}
    C_{\mvec{l}} = \frac{1}{\Omega_s} R^{\text{T}}_{\mvec{l}}\Lambda_{\mvec{l}} R_{\mvec{l}},
\end{equation}
where, $\Lambda_{\mvec{l}}$ and $R_{\mvec{l}}$ contain the eigenvalues and eigenvectors of the power spectrum matrix. Note that $C_{\mvec{l}}, \Lambda_{\mvec{l}}, R_{\mvec{l}}$ are all functions of the cosmological parameters and the wave number, $\mvec{l}$. However, for brevity, we do not explicitly show the cosmology dependence. We can now define a transformed variable, $x_{\mvec{l}}$ such that 
\begin{equation}\label{eqn:kappa_reparameterization}
    \tilde{y}_{\mvec{l}} = R^{\text{T}}_{\mvec{l}}\Lambda^{1/2}_{\mvec{l}} x_{\mvec{l}}.
\end{equation}
By construction, $x_{\mvec{l}}$ is a random variable with unit variance and therefore has no cosmological information. This reparameterization helps decorrelate the map variables from the cosmological parameters, improving sampling efficiency. For code development and testing, in addition to our lognormal model it is useful to consider a model in which $\kappa$ is described as a Gaussian random field.  We perform a similar transformation when working in the Gaussian random field model. Figure \ref{fig:fwd_model} summarizes the mathematical operators relating the sampled field $x_{\mvec{l}}$ to the convergence field in both the Gaussian and lognormal models.We adopt the shorthand $\mvec{X} = \{x_{\mvec{l}}\}$ to denote the collection of all $x$ parameters. For the rest of the paper, we perform the block sampling by sampling the $\mvec{X}$ variable instead of the $\mvec{\kappa}$ variables.

\subsubsection{Map sampling scheme}\label{sssec:map_sampling_scheme}

We want to sample $\mX$ from the posterior, $\mP(\mX|\Theta, \mD)$. Using Bayes Theorem, we can write the posterior as
\begin{equation}\label{eqn:posterior}
    \mP(\mX|\Theta, \mD) \propto \mP(\mD|\mX, \Theta) \mP(\mX|\Theta).
\end{equation}
Let us first focus on the likelihood term, $\mP(\mD|\mX, \Theta)$. The relationship between $\mX$ and the true pixelized shear field is deterministic (see Figure \ref{fig:fwd_model}). However, this relationship depends on the cosmological parameters, as $R_{\mvec{l}}$ and $\Lambda_{\mvec{l}}$ depends on the cosmological parameters. That is, we can write the shear field as a function of $\mvec{X}$ and $\Theta$ as, $\gamma^{a}_i \equiv \gamma^{a}_i(\mX, \Theta)$. Since each pixel contains a large number of galaxies\footnote{For a Stage-IV survey such as the Vera Rubin observatory, the typical number of galaxies in a (5 arcmin)$^2$ pixel will be $\sim 700$.}, the central limit theorem suggests the observed shear will exhibit Gaussian noise. Thus, given the pixelized values of the observed shear field, we can write the likelihood term for $\mX$ as,
\begin{equation}\label{eqn:likelihood}
    \mP(\mD|\mX, \Theta) \propto \prod_{j=1}^{N_{\text{bins}}}\exp \Bigg[-\sum_{i=1}^{N_{\text{pix}}}\sum_{a=1}^2 \frac{(\gamma^a_i(\mX, \Theta) - \gamma^{\text{obs},a}_i)^2}{2 \sigma^2_{\eps, i}} \Bigg],
\end{equation}
where, $\sigma_{\eps, i}$ is the shape noise per ellipticity component in the $i$-th pixel, $N_{\text{bins}}$ is the number of tomographic bins and $N_{\text{pix}}$ is the number of pixels in each redshift bin. Here, we have assumed that the noise in each pixel is independent of other pixels. 

Next, we discuss the prior, $\mP(\mX | \Theta)$. By construction, $\mP(\mX | \Theta)$ is independent of cosmology and has unit variance for all the Fourier modes. In this work we consider both a Gaussian and a lognormal model for $\mvec{\kappa}$, which impacts the deterministic relation between $\mvec{X}$ and $\mvec{\kappa}$ as described above.

We use Hamiltonian Monte Carlo \citep[HMC,][]{Neal2012} to sample the posterior distribution for $\mvec{X}$.  HMC is a sampling scheme that uses gradient information to suppress the random walk nature of traditional Monte Carlo methods, and can be used to sample from very high dimensional posteriors. To do so, HMC introduces a set of auxilliary `momentum' variables. By following the energy preserving `Hamiltonian' trajectories in the extended parameter space, the algorithm achieves very high acceptance rate. We implement our code in the Python package {\sc jax} \citep{jax2018github}, where the gradients of the functions are accessible using automatic differentiation. 

The performance of HMC is very sensitive to the choice of the `mass matrix'. It has been shown that the optimal choice of the mass matrix is the inverse of the covariance matrix of the target (posterior) distribution \citep{Taylor2008}. Because $\mvec{X}$ is sampled from a unit Gaussian distribution, the posterior for prior-dominated runs have unit variance. This can be used as an initial guess for the HMC mass matrix. We run an initial chain with this mass matrix. However, if the likelihood term dominates, the variance in the posterior will be different. Therefore, after the initial run, we  estimate our final mass matrix numerically from samples of this initial HMC chain. In this estimation, we assume that the covariance matrix of the $\mX$-modes is diagonal as per the homogeneity constraint, equation \eqref{eqn:y_cov}, and that this matrix is cosmology independent. However, we account for the correlation in wave modes across different  tomographic bins.

%%%%%%%%%%%%%%%%%%%%%%%%%%%%%%%%%%
%%%%%%%%%%%%%%%%%%%%%%%%%%%%%%%%%%
%%%%%%%%%%%%%%%%%%%%%%%%%%%%%%%%%%
%%%%%%%%%%%%%%%%%%%%%%%%%%%%%%%%%%

\subsection{Map-based cosmology sampling}\label{ssec:cosmo_sampling}

In our block sampling scheme, the cosmological parameters are sampled keeping the $\mX$ maps fixed. Using Bayes Theorem, the conditional posterior for the cosmological parameters can be written as
\begin{equation}\label{eqn:cosmo_posterior}
    \mP(\Theta | \mX, \mD) \propto \mP(\mD|\mX, \Theta) \mP(\Theta).
\end{equation}
The first term on the right hand side of this equation is the likelihood term given in equation \eqref{eqn:likelihood}. Note that while the $\mX$ fields are held fixed in this step, the convergence/shear field are not. The same $\mX$ field can result in different convergence/shear fields for different cosmological parameters since the power spectrum and the shift parameters of the convergence fields depend on cosmological parameters. We calculate the convergence power spectrum using the {\sc camb} package \citep{Lewis2000, Howlett2012}. For the lognormal prior, we also need to calculate the lognormal shift parameter. We use the {\sc cosmomentum}\footnote{\url{https://github.com/OliverFHD/CosMomentum}} \citep{Friedrich2018, Friedrich2020} code, that uses perturbation theory to calculate the cosmology-dependent shift parameters. Note that the calculation of the shift parameter in the code assumes a cylindrical window function, while for our case, the pixels are rectangular. We calculate the shift parameters at a characteristic scale, $R_{\text{cyl}} = \Delta L / \sqrt{\pi}$, where, $\Delta L$ is the resolution of the pixels. Since this code is based on perturbation theory, it may not be accurate at small-scales. However, since the main objective of this paper is to present the code and test the feasibility of this method, this is not a major issue. In the future, we will fit the shift parameters directly from $N$-body simulations. 

We use an emulator to speed up the calculation of the cosmology dependent power spectrum and shift parameters. To calculate a cosmology dependent quantity  $\mvec{D}(\Theta)$ --- e.g. power spectrum or shift parameters --- we first create a Latin hypercube sampling, $\{\Theta_i\}$ of the cosmological parameter space. We then calculate the associated data vectors, $\mvec{D}_i = \mvec{D}(\Theta_i)$ at the sampled cosmological parameters and do a Principal Component Analysis (PCA) decomposition of the calculated data vectors, 
\begin{equation}
    D_i = \sum_{j=1}^{N_{\text{PCA}}} c_{ij} Q^j.
\end{equation}
We then interpolate the PCA coefficients, $c_{ij}$ as a function of the multidimensional cosmological parameter space using a polynomical chaos expansion (PCE) emulator. We do so using the Python package, {\sc chaospy}\footnote{\url{https://chaospy.readthedocs.io/en/master/}}. PCE emulators have previously been used for cosmological applications in \citet{Knabenhans2019, Kokron2021}. After we train our emulator, we can compute the cosmology dependent power spectrum/shift parameters extremely fast.

To assess the accuracy of the emulator we perform a `leave-one out' test on our emulator. We remove one data point from the training set, retrain the emulator with the remaining data, and then test the accuracy of the emulator at the ``missing'' data point. Our emulator predicts the power spectrum (shift parameters) at $\lesssim~0.2\%$ ($0.5\%)$ accuracy, more than sufficient for our purposes.

We use an adaptive slice sampler \citep{Neal2000} to sample from the posterior of the cosmological parameters, equation \eqref{eqn:cosmo_posterior}. In this paper, we only sample the cosmological parameters, $\Omega_m, \sigma_8$. Since we impose a lognormal prior on the convergence field, our cosmological inference scheme is sensitive to the non-Gaussian information present in the lensing maps. As we will see in section \ref{sec:results}, using the lognormal prior improves cosmological parameter inference over power spectrum only inference.
\section{Code Validation with Analytic Posteriors}\label{sec:performance_test}

\begin{figure}
    \centering
    \includegraphics[width=\linewidth]{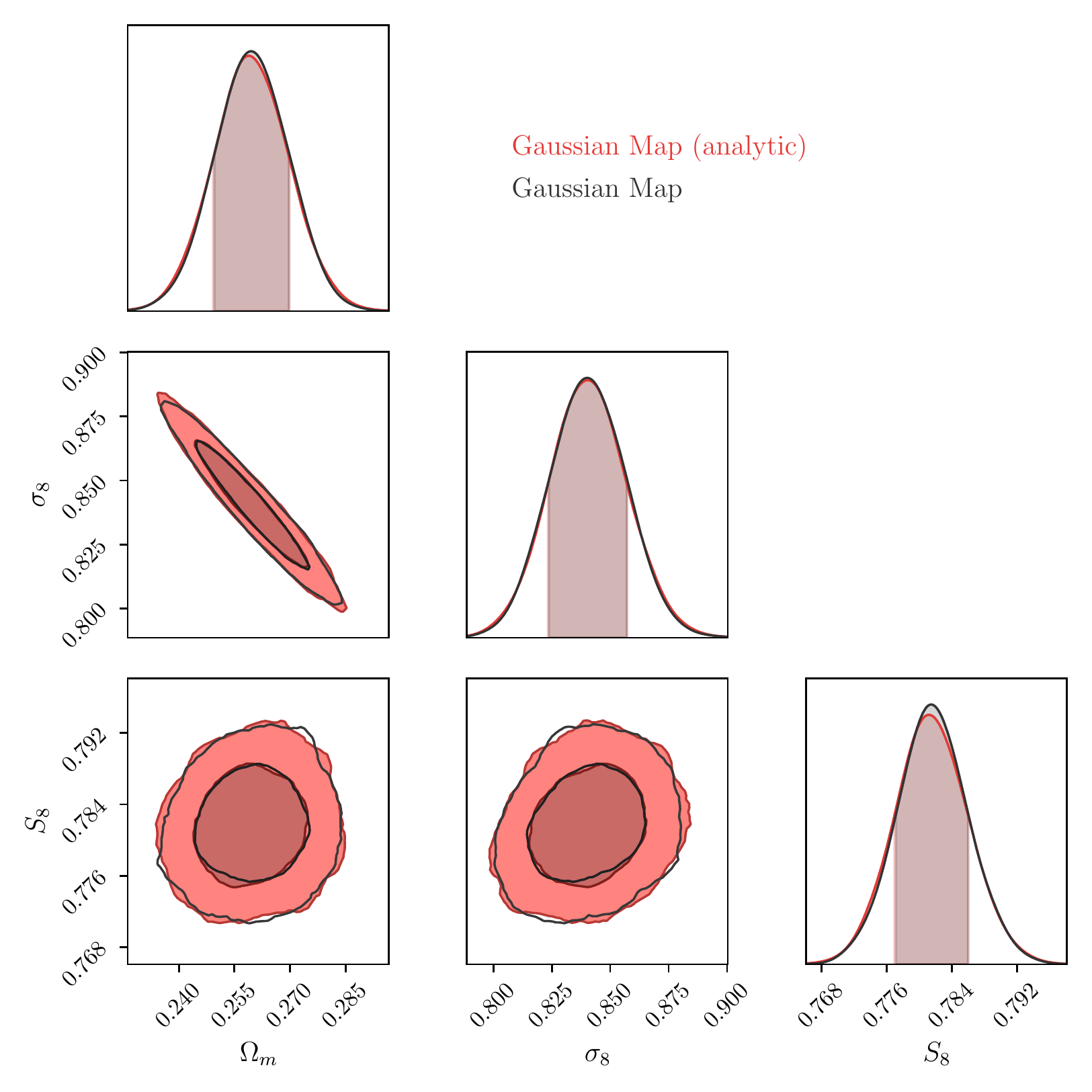}
    \caption{Comparison of the cosmology contours obtained from our map-based inference code (black contours) and the analytic marginalization of the mass maps (solid red contours) as described in section \ref{sec:performance_test}. In this comparison, we mask the Nyquist frequencies of the KS matrix, $\mmat{A}$, since these frequencies are rendered non-Hermitian due to resolution effects. The resulting contours from the HMC code and the analytical marginalization based code are remarkably similar. The standard deviation in the two parameters ($\Omega_m$ and $\sigma_8$) calculated from the samples are accurate to $< 1\%$ and the shift in the mean of $\Omega_m$ and $\sigma_8$ is less than $0.01\sigma$.}
    \label{fig:analytic_cosmo_contours}
\end{figure}

We wish to verify that our code correctly marginalizes the cosmological posteriors over the map distribution.  If the convergence field $\mvec{\kappa}$ is Gaussian, the $\mvec{\kappa}$ field can be marginalized analytically. This allows us to compare our sampling-based results to the analytic posteriors.  Let us then derive the results of analytical marginalization for a Gaussian prior. Using Bayes Theorem, we can write the posterior on the cosmological parameters, $\Theta$, as
\begin{equation}
    P(\Theta|\mD) \propto P(\mD|\Theta)P(\Theta),
\end{equation}
where, $P(\Theta)$ is the prior on the cosmological parameters. The likelihood term involves a marginalization over the convergence maps, $\mvec{\kappa}$ and can be written as, 
\begin{equation}\label{eqn:marginalized_likelihood}
    P(\mD | \Theta) = \int d\mvec{\kappa}~ P(\mD|\mvec{\kappa}) P(\mvec{\kappa}|\Theta).
\end{equation}
In the above equation, we have assumed that the data depends on $\Theta$ only through its dependence on $\mvec{\kappa}$. Assuming $\mvec{\kappa}$ is a Gaussian random field we have that
\begin{equation}\label{eqn:prior}
    P(\mvec{\mvec{\kappa}}|\Theta) = \frac{1}{\sqrt{\text{det}(2\pi \mmat{S}_{\kappa})}}\exp\bigg[-\frac{1}{2}\mvec{\kappa}^{\dag} \mmat{S}^{-1}_{\kappa} \mvec{\kappa}\bigg],
\end{equation}
where the covariance matrix of the map, $\mmat{S}_{\kappa}$, depends on the cosmological parameters. The shear field depends on the convergence field through the forward Kaiser-Squires transformation, equation \eqref{eqn:fwd_KS}. The discretized version of this transformation can be expressed as a matrix multiplication by the transformation matrix $\mmat{A}$, where
\begin{equation}
    \mvec{\gamma} = \mmat{A}\mvec{\kappa}.
\end{equation}
So far, we have not assumed any explicit basis representation of the $\mvec{\kappa}$ maps. Since the shear field is non-local, the matrix $\mvec{A}$ is dense in real space. However, in Fourier space, the transformation is diagonal and can be expressed as, 
\begin{equation}
    \tilde{\mmat{A}}_{ij} = \frac{l^2_{x,i} - l^2_{y,i} - 2 i l_{x,i}l_{y,i}}{l^2_{x,i} + l^2_{y,i}}\delta^{K}_{ij}.
\end{equation}
Assuming Gaussian shape noise, the probability of the observed shear field is
\begin{equation}\label{eqn:lkl}
    P(\mD|\mvec{\kappa}) = \frac{1}{\sqrt{\text{det} (2\pi\mmat{N})}}\exp\bigg[-\frac{1}{2}
    (\mvec{\gamma}_{\text{obs}} - \mmat{A} \mvec{\kappa})^{\dag}\mmat{N}^{-1} (\mvec{\gamma}_{\text{obs}} - \mmat{A} \mvec{\kappa}) \bigg],
\end{equation}
where $\mmat{N}$ is the covariance matrix describing the noise. Combining equations \eqref{eqn:prior} and \eqref{eqn:lkl}, we arrive at
\begin{align}\label{eqn:analytic_posterior_integrand}
 P(D|\mvec{\kappa}) P(\mvec{\kappa} | \Theta) &\propto \nonumber \\ 
 \frac{1}{\sqrt{\text{det}(2\pi \mmat{S}_{\kappa}})} &\exp \bigg[-\frac{1}{2} (\mvec{\kappa} - \mvec{\kappa}_{\text{WF}})^{\dag} \mmat{S}^{-1}_{\text{WF}}(\mvec{\kappa} - \mvec{\kappa}_{\text{WF}})\bigg] \nonumber \\  &\exp\bigg[\frac{1}{2}\mvec{\kappa}^{\dag}_{\text{WF}}\mmat{S}^{-1}_{\text{WF}} \mvec{\kappa}_{\text{WF}}\bigg],
\end{align}
where
\begin{equation}
    \mmat{S}_{\text{WF}} = (\mmat{S}^{-1}_{\kappa} + \mmat{A} \mmat{N}^{-1}\mmat{A}^{\dag})^{-1},
\end{equation}
and $\mvec{\kappa}_{\text{WF}}$ is the Wiener-filtered convergence field.  This is given by
\begin{equation}
    \mvec{\kappa}_{\text{WF}} = \mmat{W} \mvec{\gamma}_{\text{obs}}.
\end{equation}
In the above equation the Wiener filter $\mmat{W}$ is defined via
\begin{equation}
    \mmat{W} = \mmat{S}_{\kappa} [\mmat{S}_{\kappa} + \mmat{A}^{\dag}\mmat{N}\mmat{A}]^{-1} \mmat{A}^{\dag}.
\end{equation}
Plugging equation \eqref{eqn:analytic_posterior_integrand} into equation \eqref{eqn:marginalized_likelihood} we arrive at an expression for the likelihood of the data given the cosmological parameters,
\begin{equation}
    \mL(\Theta) \equiv P(\mD | \Theta) \propto \sqrt{\frac{\text{det}(2\pi \mmat{S}_{\text{WF}})}{\text{det}(2\pi \mmat{S}_{\kappa})}} \exp\bigg[\frac{1}{2}\mvec{\kappa}^{\dag}_{\text{WF}}\mmat{S}^{-1}_{\text{WF}} \mvec{\kappa}_{\text{WF}}\bigg].
\end{equation}
We can use the above equation to analytically evaluate the likelihood of the cosmological parameters given the data.  Note that to do so we must perform Wiener Filtering of the maps at each step in the sampling of the cosmological parameters. In general, there is no basis in which both the $\mmat{S}_{\kappa}$ and $\mmat{N}$ are sparse. Therefore, Wiener filtering of the convergence field is not trivial \citep[See ][]{Elsner2013}. $\mmat{S}_{\kappa}$ is diagonal in the Fourier basis, while the noise covariance, $\mmat{N}$ is diagonal in the real basis. However, we can simplify by assuming that the number of galaxies in each pixel are the same. In that case, the noise covariance matrix is diagonal in both the real and Fourier basis and we can perform fast Wiener filtering of the convergence field. 

Note that the Kaiser-Squires inversion matrix, $\mmat{A}$ becomes non-hermitian for the Nyquist frequencies due to resolution effects. Therefore, in order to compare the cosmological contours of our HMC code and the analytically-marginalized posterior, we mask the Nyquist frequencies in our simulations and in our inference. We compare the cosmological contours obtained from the HMC-based posterior and the analytical posterior for an LSST-like survey with $5$ tomographic bins (described in detail in section \ref{ssec:cosmology_inference}) and a pixel resolution of $20$ arcminutes in Figure \ref{fig:analytic_cosmo_contours}. In the HMC-based code we sample a total of $\gtrsim 46000$ dimensions, while for the analytical posterior, we sample only $2$ dimensions. As we can see from the Figure, the two contours are remarkably similar. The standard deviation in $\Omega_m$ and $\sigma_8$ as calculated using $\sim~5000$ independent samples are different by $\lesssim 1\%$. Furthermore, the differences in the mean value of $\Omega_m$ and $\sigma_8$ in the samples are $ 0.01 \sigma$ and $0.004\sigma$ respectively. The excellent agreement of the two posteriors establishes the feasibility and accuracy of our proposed map-based cosmology inference method.

\section{Results}\label{sec:results}

\subsection{Survey Assumptions}
\label{ssec:assumptions}

\begin{figure}
    \centering
    \includegraphics[width=\linewidth]{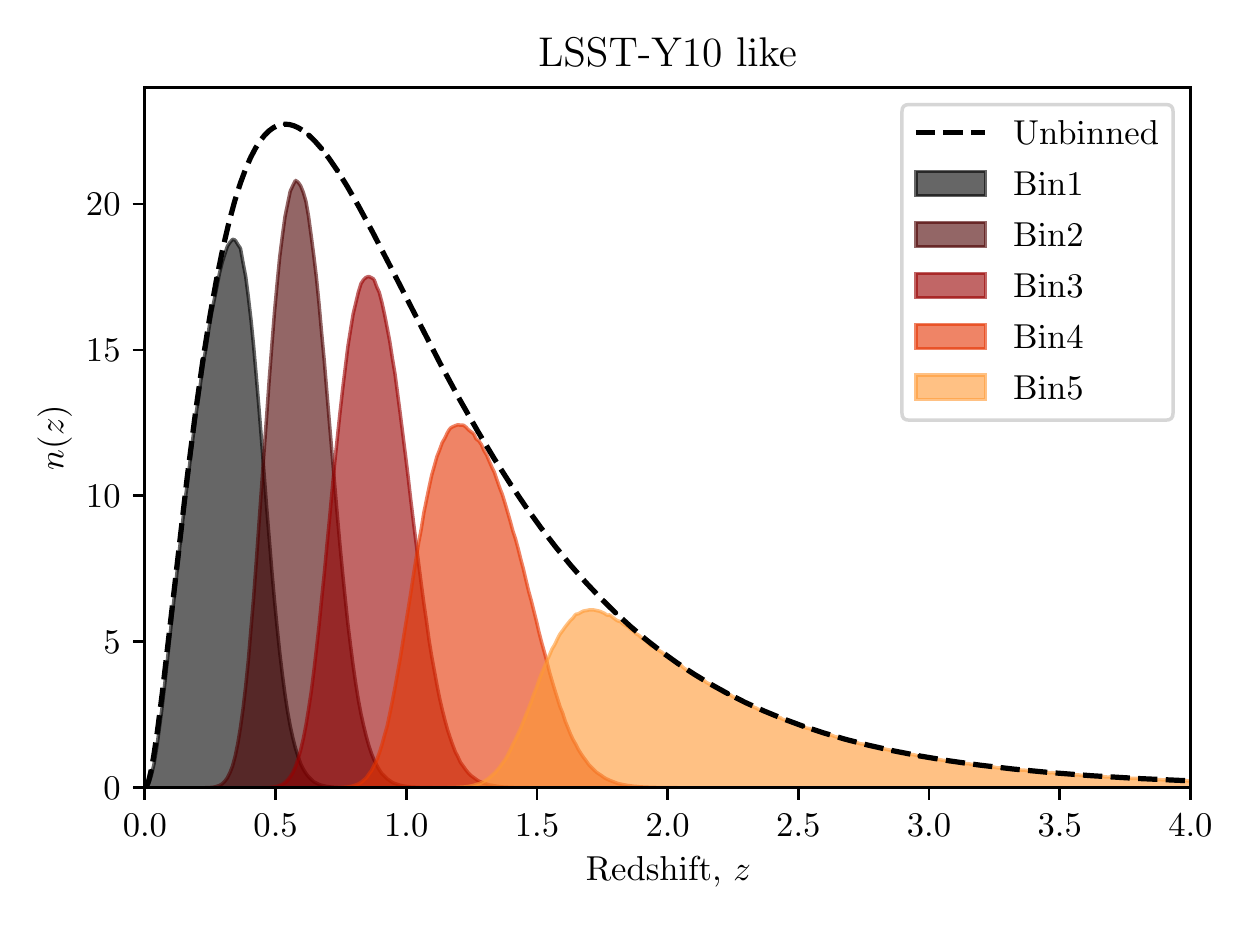}
    \caption{Redshift distribution, $n(z)$ of the our LSST-Y10 like mock survey. The redshift distribution is obtained using a Smail distribution assuming values of $(z_0, \alpha) = (0.11, 0.68)$. The tomographic bins are created assuming that each redshift bin consists of equal number of galaxies. The assumed photo-$z$ error is $0.05(1+ z)$. In this plot, we choose a source number density of $30$ arcmin$^{-2}$.}
    \label{fig:n_z}
\end{figure}

We characterize the performance of our map-based inference algorithm at a variety of source densities and resolutions.  We assume a redshift distribution similar to that expected for the LSST-Y10 data set.  Specifically, the source redshift distribution is parameterized using the Smail distribution \citep{Smail1995} as
\begin{equation}
    n(z) \propto z^2 \exp [ -(z/z_0)^{\alpha}],
\end{equation}
where the values of $z_0$ and $\alpha$ used are: $(z_0, \alpha) = (0.11, 0.68)$ for LSST-Y10 \citep{LSST_SRD}. We assume a photometric redshift error of $\sigma_z = 0.005 (1 +z)$, and create mock catalogues assuming three different source number densities: $10$, $30$ and $50$ arcmin$^{-2}$.  These correspond roughly to DES-Y6 like, LSST-Y10 like, and Roman Observatory like source densities respectively.  The sources are then split into 5 tomographic bins with equal numbers of galaxies. The resulting source redshift distribution is shown in Figure \ref{fig:n_z}. All of our mock surveys have an area of ($32$ deg)$^{2}\approx 1000\ {\rm deg}^2$.  We characterize the impact of resolution on our results by considering maps with three different resolutions, corresponding to pixel sizes of 5, 10, and 20 arcminutes.

For this paper, we perform our inferences on lognormal mocks. We generate our lognormal mocks using the power spectrum calculated assuming the above redshift distribution and tomographic binning. The shift parameter used in the generation of the mocks is calculated using the {\sc cosmomentum} code.

\subsection{Tomographic mass map reconstruction}\label{ssec:tomo_mass_map}

\begin{figure*}
    \centering
    \includegraphics[width=\linewidth]{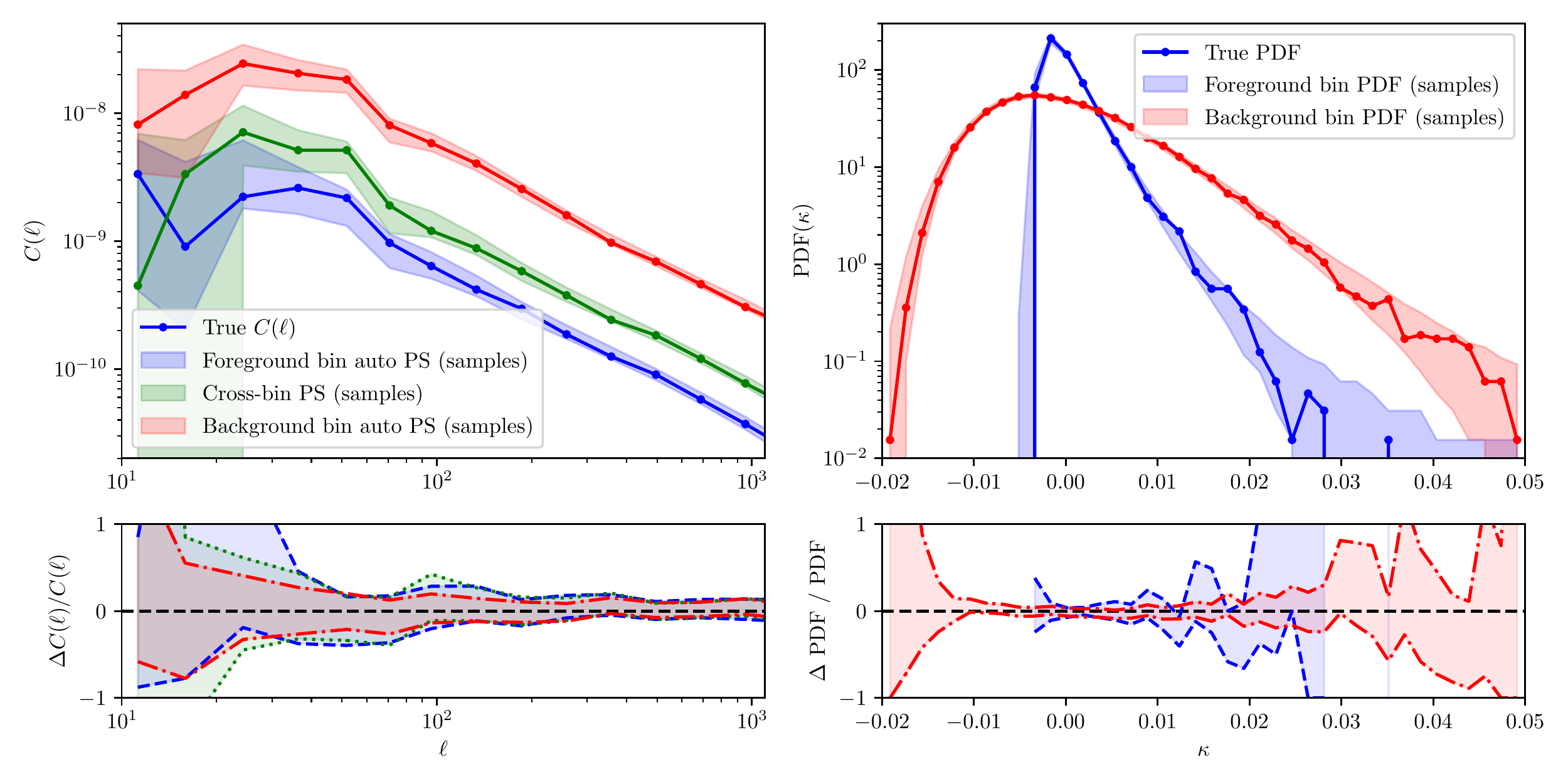}
    \caption{Summary statistics of the reconstructed maps and comparison with the true maps. \textit{(Top left):} The true binned power spectrum of the mock data shown with solid lines and dots. The blue and the red curves show the auto power spectrum for the foreground and background bins respectively, while the green curve shows the cross power spectrum between the two bins. \textit{(Top right):} The one point PDF of the sampled maps compared to the true one point PDF. \textit{(Bottom)}: The residual plots for the power spectra (\textit{left}) and the one point function (\textit{right}). The shaded regions denote the 95\% confidence intervals in the power spectrum and the PDF calculated from the posterior map samples. As can be seen, our algorithm recovers both the one point and two point statistics in an unbiased manner.}
    \label{fig:summary_stats}
\end{figure*}

\begin{figure*}
    \centering
    \includegraphics[width=\linewidth]{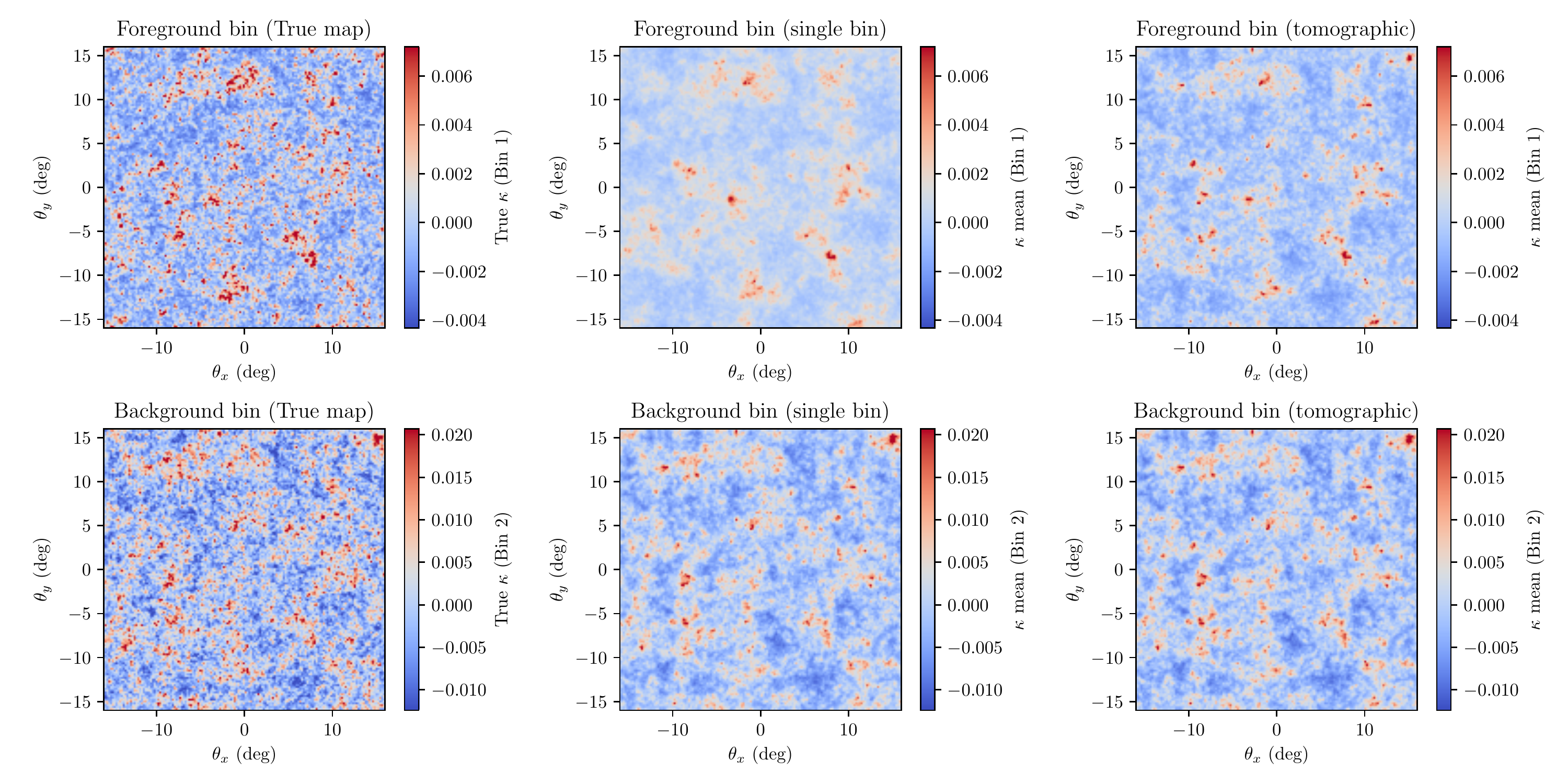}
    \caption{{\it Left:} The true convergence maps. {\it Centre:} the mean maps obtained using single bin reconstruction. {\it Right:} the mean maps obtained from a tomographic reconstruction that properly accounts for the covariance between the bins. The top row shows the maps for the foreground bin (bin 1 of the LSST-Y10 like redshift distribution) and the bottom row show the maps for the background bin (bin 3 of the LSST-Y10 like redshift distribution). The small scale structures in the maps are smoothed out due to the presence of shape noise. This is most prominent for the foreground bin in the single bin reconstruction. However, notice that smaller structures are recovered when the cross-correlations are used in the reconstruction, as can be seen by comparing the right and centre panels of the top row.
    }
    \label{fig:tomo_mass_map}
\end{figure*}

\begin{figure}
    \centering
    \includegraphics[width=\linewidth]{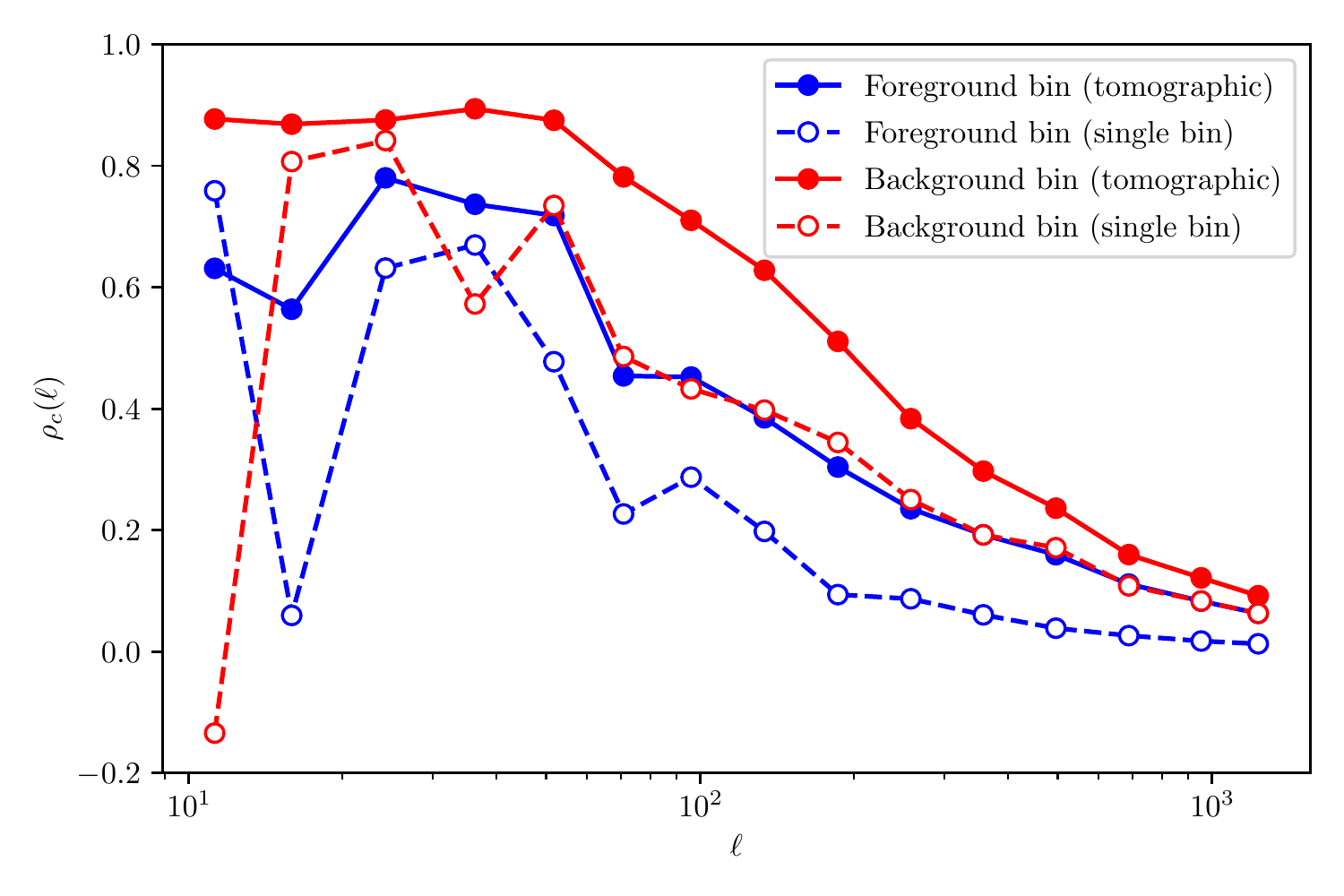}
    \caption{The mean cross-correlation of the reconstructed maps with the true convergence maps for the single-bin (dashed line) and joint-bin (solid line) reconstructions as a function of $l$. The foreground and the background bins are shown with blue and red colors respectively. We see that for both redshift bins, the correlation with the truth is substantially improved when the covariance between the two redshift bins is considered.
    }
    \label{fig:tomo_cross_corr}
\end{figure}

We first illustrate the improvements in mass mapping obtained by including cross-correlation information between different redshift bins. To do so, we jointly constrain the mass maps for two redshift bins:  bins 1 and 3 of our LSST-Y10 like redshift distribution defined in section \ref{ssec:assumptions}. In this section, we use a DES-Y6 like source density of 10~arcmin$^{-2}$ , corresponding to source density of 2~arcmin$^{-2}$ per bin. We adopt a pixel resolution of $10$~arcminutes. Figure \ref{fig:summary_stats} illustrates the fidelity of our two mass maps. We find the 1- and 2-point functions recovered from our posterior map distributions are unbiased relative to truth.\footnote{For a comparison to simulations and further discussion, see \citet{Fiedorowicz2022}.} Moreover, we successfully recover the cross power spectrum between the two tomographic redshift bins. This last success is only possible because we explicitly account for the covariance between the two mass maps, a condition that is often ignored in the literature.\footnote{A notable exception is the mass mapping method of \citet{Alsing2016, Alsing2017}}

To further demonstrate the importance of taking into account the covariance between tomographic bins we make 3 runs of our code as follows: {\it i)} foreground bin alone (Bin 1 in the LSST-Y10 tomographic binning); {\it ii)} background bin alone (Bin 3 in the LSST-Y10 tomographic binning); and {\it iii)} both bins together, accounting for the cross-correlations in the sampled maps. In all of the runs, we simultaneously sample the cosmological parameters.

Figure \ref{fig:tomo_mass_map} shows the true convergence maps along with the mean maps obtained from the three different runs. Note that in the mean maps, the small-scale structures are smoothed out due to shape noise. This is similar to the Wiener filtering of the maps. Comparing our single bin reconstructions of the foreground and background bins (middle column in Figure~\ref{fig:tomo_mass_map}), it is clear that despite the two bins having the same number of source galaxies, small scale structure is better resolved in our high redshift bin. This is because the more distance bin has a stronger lensing signal, and therefore higher signal-to-noise. The third column in Figure~\ref{fig:tomo_mass_map} shows the mean mass maps obtained when simultaneously sampling both tomographic bins.  Comparing  these to the maps in the second column (i.e. the single bin reconstructions), we see that the small scale structure is better resolved when we jointly sample both tomographic bins.  The improvement is particularly pronounced in the case of the foreground bin.  

Figure~\ref{fig:tomo_cross_corr} formalizes our visual impression, demonstrating that including the cross-correlations in the mass mapping procedure results in substantially improved cross-correlations between the sampled maps and the true maps.  That is, accounting for the covariance across tomographic bins improves the fidelity of the reconstructed mass maps.

%%%%%%%%%%%%%%%%%%%%%%%%%%%%%%%%
%%%%%%%%%%%%%%%%%%%%%%%%%%%%%%%%
%%%%%%%%%%%%%%%%%%%%%%%%%%%%%%%%
%%%%%%%%%%%%%%%%%%%%%%%%%%%%%%%%
%%%%%%%%%%%%%%%%%%%%%%%%%%%%%%%%

\subsection{Importance of Marginalizing Cosmology in Mass Mapping}
\label{ssec:cosmology_marg}
\begin{figure}
    \centering
    \includegraphics[width=\linewidth]{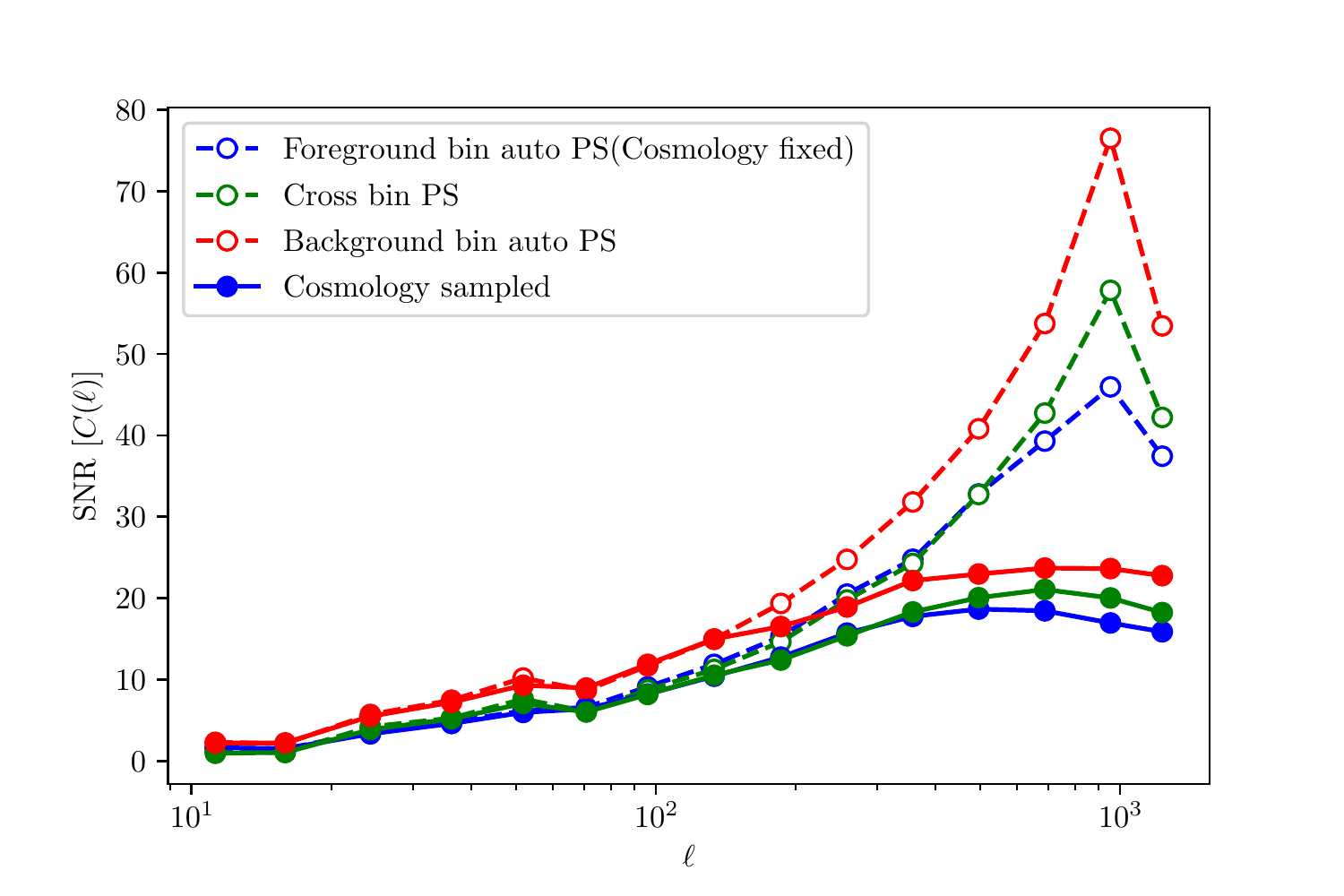}
    \caption{The signal-to-noise ratio (SNR) of $C(\ell)$, defined as the ratio of the mean estimate of $C(\ell)$ to its standard deviation for the two auto and one cross power spectrum of the two tomographic bins considered. The blue and the red curves show the auto power spectrum for     the foreground and background bin respectively. The green curve is for the cross-power spectrum between the two bins. The solid curves are the signal-to-noise when marginalizing over cosmology, while the dashed curves show the SNR with cosmology held fixed. We see the uncertainty in the power spectrum of the large $\ell$ modes (small scales) is underestimated (the SNR therefore is overestimated) if cosmological parameters are fixed.
    }
    \label{fig:Cl_SNR}
\end{figure}

The original \karmma\ algorithm assumed a fixed cosmology when performing mass map reconstruction.  This is an important drawback: fixing the cosmological parameters imposes a tight prior on the convergence maps, especially for the small scale modes that are prior-dominated. Since we do not in fact know the cosmological parameters a priori, this prior leads to unrealistically tight posterior distributions for the mass maps. Thus, to capture the true uncertainty in the mass maps we need to simultaneously vary the cosmological parameters.

We demonstrate the impact of cosmology marginalization on the mass maps for the same two sample bins from the previous section. Figure \ref{fig:Cl_SNR} shows the signal-to-noise (SNR) ratio of the inferred power spectrum as a function of $\ell$ in two scenarios. In the first scenario, we sample the cosmological parameters simultaneously with the mass maps, while in the second scenario, we keep the cosmological parameters fixed to their fiducial value. We define the SNR ratio as
\begin{equation}
    \text{SNR}[C(\ell)] = \frac{\text{Mean}[C^{\text{samples}}(\ell)]}{\text{Std. dev}[C^{\text{samples}}(\ell)]},
\end{equation}
where $C^{\text{samples}}(\ell)$ is the binned power spectrum measured from the reconstructed mass map samples. As we see from the figure, the SNR ratio at low $\ell$ (large-scales) is minimally impacted by the simultaneous sampling of the cosmological parameters. This reflects the fact that the large scale modes are likelihood dominated. On the other hand, the SNR of the small scale (large $\ell$) modes is substantially overestimated if the cosmological parameters are held fixed. Underestimating the uncertainty of these modes may impact, e.g, the assessment of significance of cross-correlations with other cosmological probes \citep[e.g, ][]{Nguyen2020}. These results demonstrates that accurately capturing the full posterior distribution of our mass maps necessitates marginalizing over the unknown cosmological parameters.  Conversely, improvements in our knowledge of the background cosmology from external probes (e.g. CMB) can further improve the fidelity of mass map reconstruction.

%%%%%%%%%%%%%%%%%%%%%%%%%%%%
%%%%%%%%%%%%%%%%%%%%%%%%%%%%
%%%%%%%%%%%%%%%%%%%%%%%%%%%%
%%%%%%%%%%%%%%%%%%%%%%%%%%%%
%%%%%%%%%%%%%%%%%%%%%%%%%%%%

\subsection{Map-based cosmological parameter inference}\label{ssec:cosmology_inference}

We now use the 5 tomographic bin set-up discussed in section \ref{ssec:assumptions} to test our map-based cosmology inference pipeline.  We will run chains with 3 different number densities ($10/30/50$ arcmin$^{-2}$) and 3 different pixel resolutions ($20/10/5$ arcminutes) to assess the impact of these parameters on our results. Specifically, we create $25$ lognormal mock realizations for each of the $3$ source number densities (10/30/50 arcmin$^{-2}$) and the $3$ pixel resolutions (20/10/5 arcmin). We use these mocks in section \ref{sssec:biases} to characterize the bias in the inferred cosmological parameters, while section~\ref{sssec:gains} discusses the improvement in our posteriors relative to standard two-point analyses.

\subsubsection{Posterior Validation}
\label{sssec:posterior_validation}

\begin{figure}
    \centering
    \includegraphics[width=\linewidth]{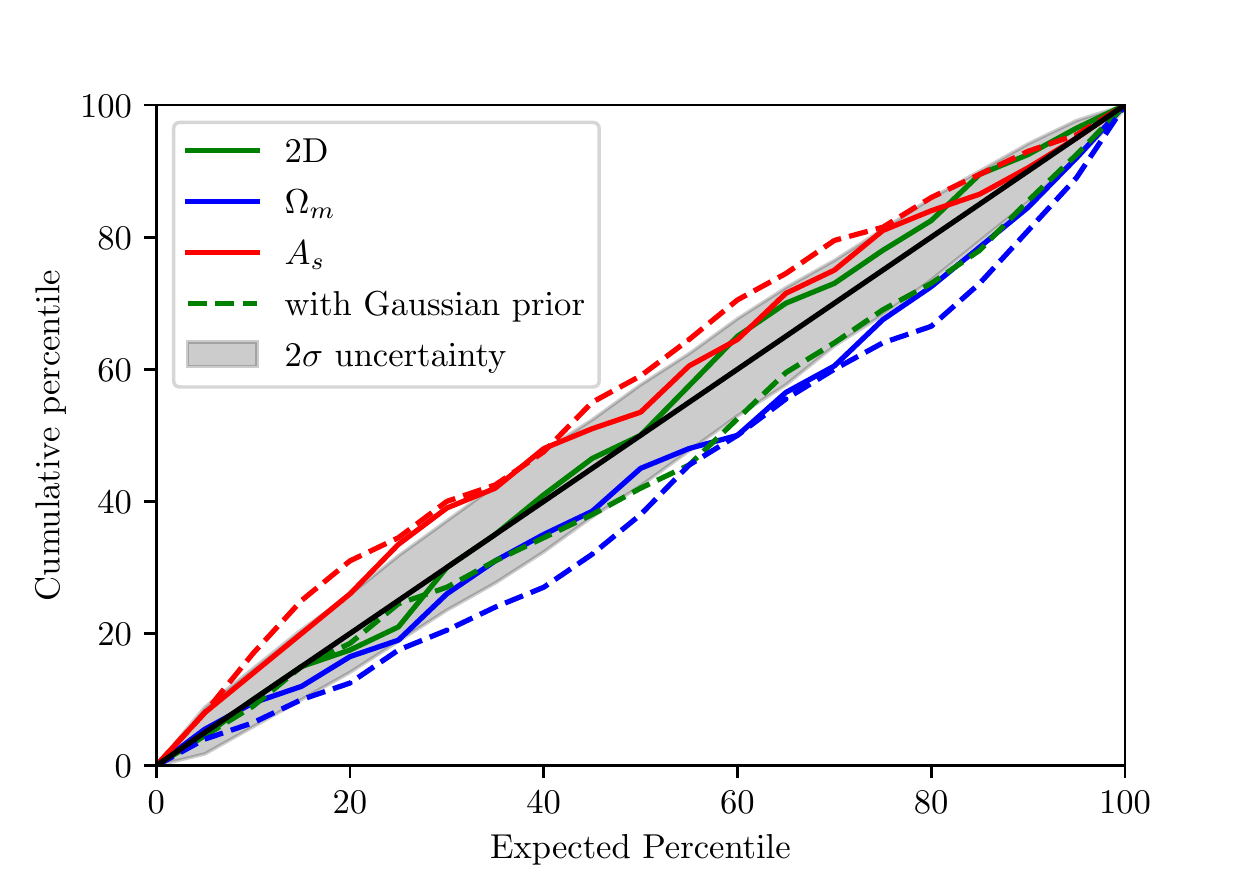}
    \caption{\pp plot for cosmological parameter inference with $200$ mock lognormal realizations. The black line and the grey shaded region is the expected mean and the $95\%$ confidence interval calculated using $5000$ Gaussian realizations. The solid blue, red and green lines show the \pp plot calculated for $\Omega_m$, $A_s$ and the 2D contours assuming a lognormal prior for the inference. The dashed lines are the corresponding \pp plots assuming a Gaussian prior for the inference.}
    \label{fig:percentile_plot}
\end{figure}

We begin our discussion of the map-based cosmology inference by verifying that our cosmological posteriors adequately characterize the uncertainty in the inferred cosmological parameters. To do so, we produce $200$ lognormal realizations of our mock survey with a source density of $30$ arcmin$^{-2}$ and $20$ arcmin pixel resolution. For each of these $200$ realizations, we run our map-based inference code assuming a Gaussian and a lognormal prior. Each run yields cosmological parameters sampled from the cosmology posterior.

Consider a confidence contour containing $X\%$ of the posterior.  If our posteriors are correct, then for $X\%$ of our mocks we should find that the true cosmological parameters are contained within the $X\%$ posterior estimated from our chains.  To verify this is indeed the case, we calculated the fraction of mocks for which the true cosmological parameters are contained within a given confidence contour along a grid of percentiles, and plot the resulting fraction as a function of the input values.  If our posteriors are correct, the resulting plot should trace the $y=x$ line. This plot is sometimes referred to as a \pp plot \citep[e.g,][]{Gibbons2011}.  Unfortunately, because each run requires running a full MCMC, we are limited to relatively few (200) mocks when applying this test.

We rely on Gaussian simulations to quantify the uncertainty in the \pp plot. We draw the mean of $200$ Gaussian distributions from a unit normal distribution. Given the mean and the standard deviation, we can calculate the percentile value of the true mean. This is similar to the $200$ realizations of the lognormal mocks used for map-based inference. We make $5000$ sets of such realizations to quantify the uncertainty in the \pp curve computed from the set of $200$ Gaussian simulations. We can see from Figure \ref{fig:percentile_plot} that for the case of lognormal prior, the resulting \pp plot is within the $95\%$ confidence interval at all percentiles. On the other hand, the \pp plots in which we use a Gaussian prior to analyze log-normal simulations clearly reach outside the 95\% confidence level.  This demonstrates that an incorrect prior can bias the resulting posteriors (see below for futher discussion).

\subsubsection{Cosmological Biases and Log-Normality}\label{sssec:biases}

\begin{figure*}
    \centering
    \includegraphics[width=\linewidth]{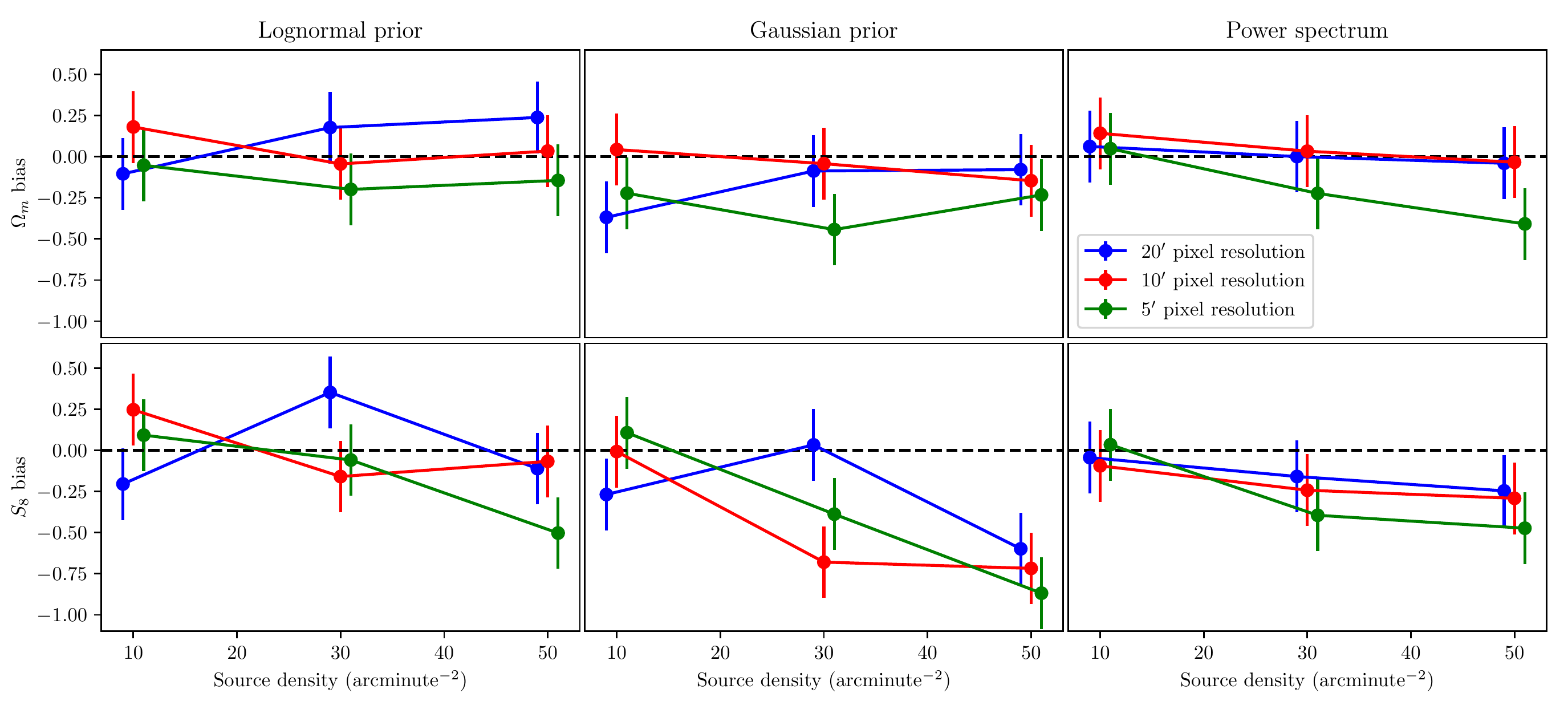}
    \caption{The bias in the inferred value of $\Omega_m$ ({\it top}) and $S_8$ ({\it bottom}) using a lognormal map-based inference ({\it left}), Gaussian map-based inference ({\it center}) and power spectrum based inference ({\it right}). The bias is defined in equation \eqref{eqn:bias}, and is calculated in units of the statistical uncertainty of the posterior. We see no significant bias in the inferred value of either $\Omega_m$ or $S_8$ for the case of lognormal map-based inference. On the other hand,
    using a Gaussian prior can lead to large ($\sim 0.8\sigma$) biases in the cosmological posteriors derived from analyzing a log-normal simulation. This demonstrates that map-based inference can lead to biased inference if the field-level prior is incorrect.
    }
    \label{fig:cosmo_bias}
\end{figure*}

To quantify the bias in a parameter $\Theta$ we compute the quantity, 
\begin{equation}\label{eqn:bias}
    B_{\Theta} = \frac{\langle \Theta \rangle - \Theta_{\text{true}}}{\sigma_{\Theta}},
\end{equation}
where, $\langle \Theta \rangle$ is the sample mean and $\sigma_{\Theta}$ is the standard deviation computed from the cosmological parameter samples. If the inferred cosmological parameters are unbiased, we will obtain a value of $B_{\Theta}$ consistent with zero. In Figure \ref{fig:cosmo_bias}, we plot the bias in the parameters $\Omega_m$ and $S_8$ as a function of the source number density and the pixel resolution. We compute the bias $B_{\Theta}$ for the $25$ realizations for each configuration of number density and pixel resolution. The uncertainty in the bias estimates are calculated from the scatter in the bias across the 25 mock realizations. The different columns in Figure \ref{fig:cosmo_bias} correspond to map-based inference using a lognormal prior (left panel), a Gaussian prior (central panel), and power spectrum inference (right panel, discussed in section \ref{sssec:gains}). When using the lognormal map-based inference, we do not see any significant bias for any of the survey parameters considered. On the other hand, when using a Gaussian map-based inference, we can see that the inferred value of $S_8$ is biased low. The maximum bias is seen at the highest source density and resolution we considered, $\bar{n} = 50$ arcminute$^{-2}$ and a pixel resolution of $5$ arcminute.  For this scenario, we find the Gaussian map based inference of $S_8$ is biased low by $\sim 0.8 \sigma$.  The statistical significance of this detection is $\sim 4\sigma$. Our results demonstrate that an incorrect prior can bias the inferred cosmological parameters. Consequently, it is possible our log-normal approach is itself biased when applied to simulation-based maps.  We postpone an investigation of these possible biases to future work, noting only that the log-normal model has been shown to accurately reproduce a variety of mass map clustering statistics.

We also notice a small bias ($\sim 0.4\sigma$) when we infer the cosmological parameters using power spectrum. This is similar to the results obtained by \citet{Leclercq2021} who found that a likelihood-based power spectrum analysis of lognormal maps leads to biases in the inferred cosmological parameters.

\subsubsection{Comparison to Standard Methods}
\label{sssec:gains}

\begin{figure*}
    \centering
    \includegraphics[width=0.8\linewidth]{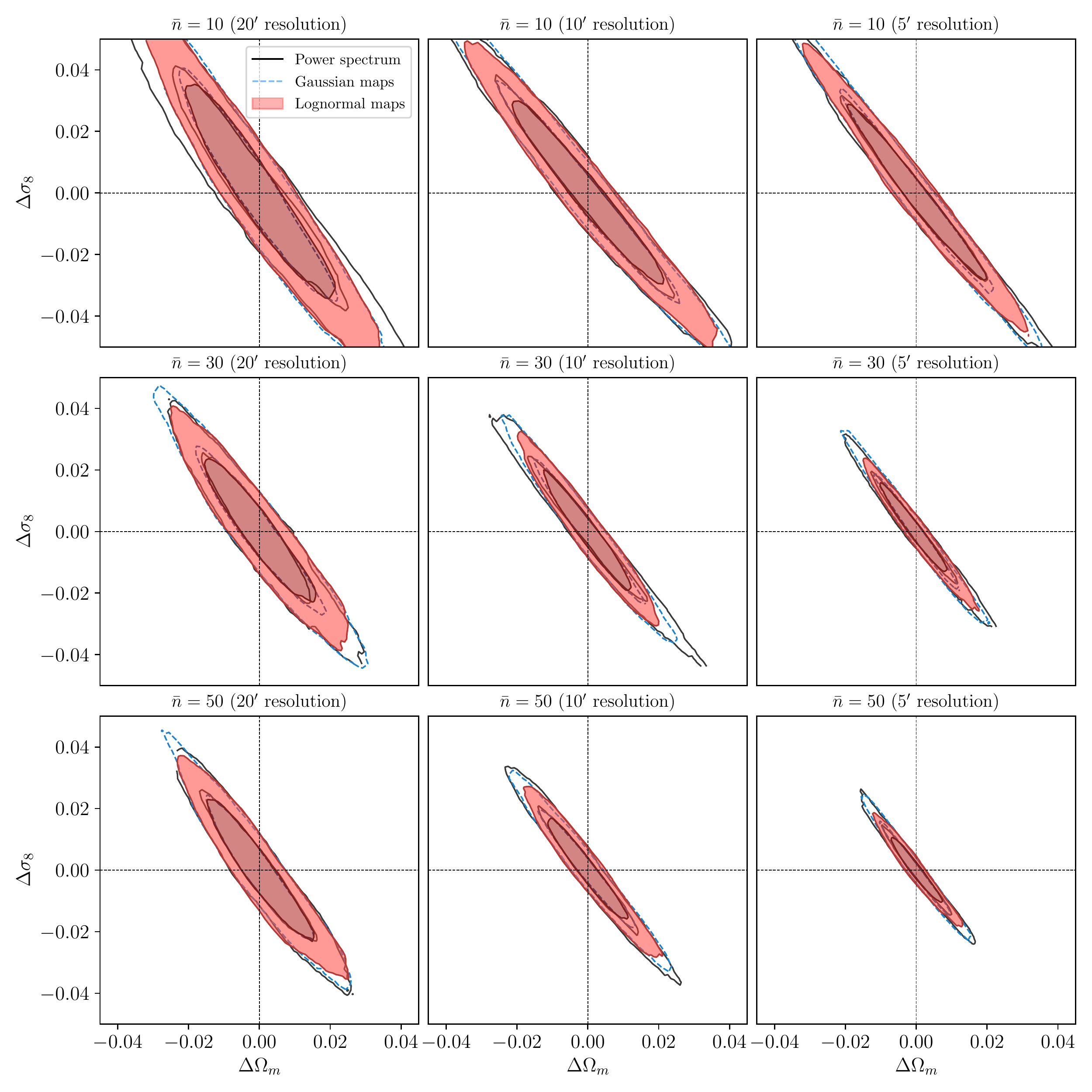}
    \caption{Cosmology contours obtained for a set of mock lognormal realizations. For easy visualization, we subtract the mean from each of the cosmological contours so that the contours are centred on $(0, 0)$. The three columns correspond to three different pixel reolutions: $20^{\prime}$ ({\it left}), $10^{\prime}$ ({\it center}) and $5^{\prime}$ ({\it right}) and the three rows correspond to the three different source number densities: $\bar{n} = 10$ arcmin$^{-2}$ ({\it top}), $\bar{n} = 30$ arcmin$^{-2}$ ({\it middle}) and $\bar{n} = 50$ arcmin$^{-2}$ ({\it bottom}). The black line shows the result of power spectrum inference, red contours corresponds to the map-based inference assuming a lognormal prior, the blue dashed lines are the results from the map-based inference assuming a Gaussian prior. We immediately notice that the lognormal contours are more constrained than the power spectrum or the Gaussian map based contours. On the other hand, the power spectrum and the Gaussian map-based cosmology contours are of similar size.}
    \label{fig:cosmo_contours}
\end{figure*}

\begin{figure*}
    \centering
    \includegraphics[width=\linewidth]{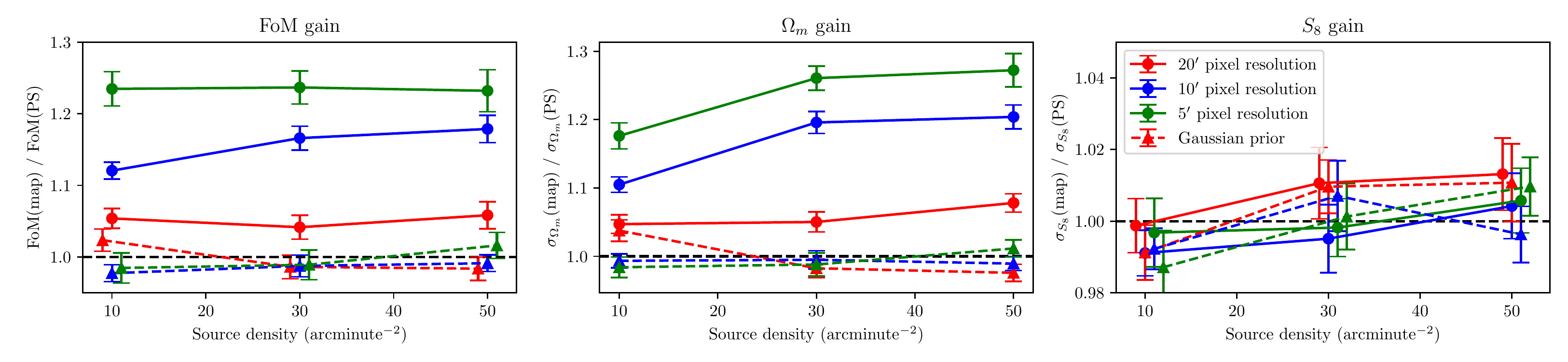}
    \caption{The gain in cosmological parameter constraints relative to power spectrum based constraints using lognormal map-based inference ({\it solid lines, circular marker}) and Gaussian map-based inference ({\it dashed lines, triangular marker}). The three colours corresponds to three different pixel resolutions: 20 arcminute ({\it red}), 10 arcminute ({\it blue}) and 5 arcminute ({\it green}). The left panel shows the gain in the FoM, the center panel shows the relative gain in $\Omega_m$ constraints and the right panel shows the relative gain in $S_8$ constraints. We see no relative gain for the Gaussian map-based inference over power spectrum only inference. On the other hand, for lognormal map-based inference, we see upto $30\%$ gain in the FoM and the $\Omega_m$ constraints. However, we see no substantial gain in the inferred value of $S_8$ using a lognormal map-based inference approach. As expected the gain in cosmological parameter inference with the lognormal map-based inference increases with higher number density of sources and smaller scales.}
    \label{fig:cosmo_gains}
\end{figure*}

We now characterize the improvement in the cosmological parameters constraints derived using the lognormal map-based inference over Gaussian map-based inference and power spectrum only inference. To perform power spectrum only inference, we use a Gaussian likelihood calculated with the Gaussian covariance of the binned power spectrum.\footnote{ We also tested our power spectrum inference using a numerical covariance calculated from lognormal mocks and using the multivariate student-$t$ likelihood of \citet{Sellentin2016}. We do not see any significant difference between the inference using the analytic covariance and the numerical covariance. Therefore, we only report the results with the analytic Gaussian covariance.} It has been shown that for future stage-IV surveys, the super-survey part of the non-Gaussian power spectrum covariance will play an important role in power spectrum based inference, while the connected trispectrum contribution will likely be negligible \citep{Barreira2018}. Note that for our map based inference, it is straightforward to include the super sample effects by including the super-survey mode in the inference. However, in this work, we do not include the super-survey modes in our inference.

Figure \ref{fig:cosmo_contours} illustrates our results. For ease of visualization, we center all contours at zero by subtracting the mean value of the parameters. We notice immediately that the contours obtained with a lognormal prior are tighter than those obtained using either the power spectrum or the Gaussian map based inference. This becomes especially clear as we go to higher resolution (5 arcminute resolution) and higher source number density ($50$ arcminute$^{-2}$). On the other hand, the Gaussian map-based inference and the power spectrum based contours are of similar size. Furthermore, we see improvement in the constraints along the $S_8$ degeneracy direction.

To quantify the gain in the constraining power of the map based inference, we compute three quantities: 
\begin{enumerate}
    \item The figure-of-merit (FoM) in the $\Omega_m$-$\sigma_8$ plane defined as, 
\begin{equation}
    \text{FoM}(\sigma_8\text{-}\Omega_m) = \text{det[Cov}(\sigma_8,\Omega_m)]^{-1/2},
\end{equation} 
    \item The standard deviation in $\Omega_m$, 
    \item The standard deviation in $S_8 = \sigma_8 \sqrt{\Omega_m / 0.3}$. 
\end{enumerate}
We then compute the ratio of the map-based inference methods (both using Gaussian and lognormal priors) to the power-spectrum inference for each of these three quantities. Our results are shown in Figure \ref{fig:cosmo_gains}. We see that the ratio of the Gaussian map-based inference to the power spectrum inference is consistent with $1$ for all number densities and pixel resolutions. On the other hand, assuming a lognormal prior can lead to substantial gains in the cosmological parameter constraints. We find up to $\sim 30\%$ improvement in the FoM and the standard deviation of $\Omega_m$ when using lognormal map-based inference. On the other hand, we do not find any significant gain in the constraints on $S_8$. As expected the gain in cosmological constraints with the lognormal map-based inference method increases as we go to smaller scales (higher resolution) and higher number density. 

The gains we find from lognormal field-level inference is modest compared to other results in the literature. For example, \citet{Ribli2019} found $\sim 2$-$3$ times improvement in cosmological constraints using convolutional neural netwrorks for an LSST like survey. Other results found similar results using non-Gaussian statistics \citep[e.g,][]{Cheng2020}. However, these results consider scales much smaller than the ones considered in this work. Furthermore, while the lognormal model is able to capture the one-point and two-point statistics of the convergence field, the full non-Gaussian information content has not been studied comprehensively to the best of our knowledge.\footnote{Note however that the one-point function is sensitive to the full hierarchy of $N$-point functions.} A field-level approach that considers similar scales as us is that of~ \citet{Porqueres2022} who find that their field-level inference improves the cosmological constraining power by factors of a few, many times more than our results. We have verified that this is driven primarily by different assumptions in the calculation of the power spectrum covariance. The two covariances differ in two ways: {\it} We include the off-diagonal terms in the covariance which is discarded by \citet{Porqueres2022}. {\it ii)} The diagonal terms differ in their shape noise. We find that the difference in the constraint is driven by a combination of the two factors, with the latter being the dominant effect. A detailed comparison of the two methods will be performed in a future work.
\section{Conclusion}\label{sec:conclusion}

In this paper, we introduced a new map-based cosmology inference method for constraining cosmological parameters from weak lensing survey data. In our method, we forward model the weak lensing data at the field level. Many such forward-modelled frameworks have been proposed in the literature. For example, \citet{Bohm2017} used a 3D lognormal density field to forward model the weak lensing data. Another promising 3D forward modelling framework proposed in \citet{Porqueres2021, Porqueres2022} uses the initial condition reconstruction code {\sc borg} to forward model the weak lensing data. The 3D reconstruction from the Gaussian  initial conditions provides a unique way to capture the non-Gaussianity introduced by the structure formation process. The method presented here reconstructs the 2D convergence field. Forward-modelled reconstruction methods for 2D convergence field reconstruction has previously been presented in \citet{Alsing2016, Alsing2017}. However, unlike the method presented here, their method does not capture the non-Gaussian information in the convergence field. The 2D reconstruction methods are faster than simulation-based methods and provide a complementary way of performing field level inference. Furthermore, it is easier to combine 2D map-level approaches with other cosmological probes such as CMB lensing.

Our method reconstructs the mass maps assuming that the convergence field can be approximated as a multivariate lognormal distribution, thus recovering the correct one point and two point statistics, as well as peak and void counts, as measured in simulations \citep{Fiedorowicz2022}. Our method also consistently accounts for cross correlations between different tomographic bins. This leads to improvements in the reconstruction of the mass maps as illustrated in section \ref{ssec:tomo_mass_map}. The sensitivity to the one point function means that our method accounts for non-Gaussian information in the cosmology inference, resulting in tighter constraints on cosmological parameters relative to two-point function methods. We used mock lognormal simulations resembling the source galaxy distributions of LSST-Y10 to forecast the bias and gain in cosmological parameter constraints from using our map based inference framework as compared to power spectrum only inference. As expected, the gain in constraining power increases at higher resolutions. We find that at $5$ arcminute resolution, the lognormal map-based inference can  result in up to $30\%$ improvement in the cosmological constraints.

While the lognormal model is an excellent approximation for the field level distribution of convergence at modest scales, it breaks down on small scales. As we saw in section \ref{sssec:biases}, incorrect priors can lead to biases in cosmological parameter inference. Therefore, improved modeling of small scale structure is desirable. One can use machine learning based generative models such as generative adversarial networks \citep[GANs,][]{Mustafa2019, Perraudin2020}, neural score matching \citep{Remy2020, Remy2022} or normalizing flows \citep{Dai2022} to learn the full probability distribution at small scales. Other avenues to improve the small scale reconstruction may involve combining the lognormal prior with a sparsity prior for the mass maps \citep{Lanusse2016, Starck2021} or Gaussianizing transforms \citep{Hall2018}.

In this work, we worked entirely in the flat sky approximation. However, the future surveys considered here are wide surveys which will require curved sky analysis. We will extend the code to work on the curved sky in a future work. The first steps towards a curved sky implementation of mass mapping using a lognormal prior were already presented in \citet{Fiedorowicz2022}. Furthermore, we performed our inference assuming no systematic effects are present. However, weak lensing data sets may be substantially impacted by various systematic effects such as intrinsic alignment, baryonic effects, photometric redshift bias, etc.  Map-level approaches can provide a unique way to model systematics including spatial variations \citep{Cragg2022} and using the full field level information \citep[e.g,][]{Tsaprazi2021, HarnoisDeraps2021}. In the future, we will extend our modelling to include these systematics to enable application of our method to existing and future data sets.

\section*{Acknowledgement}
We thank Natalia Porqueres for useful discussions. The computation presented here was performed on the High Performance Computing (HPC) resources supported by the University of Arizona TRIF, UITS, and Research, Innovation, and Impact (RII) and maintained by the UArizona Research Technologies department.  SSB is supported by the DOE Cosmic Frontier program, grant DE-SC0020215. ER is supported by DOE grant DE-SC0009913 and NSF grant 2009401.  PF is supported by NSF grant 2009401.

\section*{Data availability statement}
The data products generated in this work will be shared on reasonable request to the authors.

\bibliographystyle{mnras}
\bibliography{field_level_lensing} % if your bibtex file is called example.bib

\begin{thebibliography}{}
\makeatletter
\relax
\def\mn@urlcharsother{\let\do\@makeother \do\$\do\&\do\#\do\^\do\_\do\%\do\~}
\def\mn@doi{\begingroup\mn@urlcharsother \@ifnextchar [ {\mn@doi@}
  {\mn@doi@[]}}
\def\mn@doi@[#1]#2{\def\@tempa{#1}\ifx\@tempa\@empty \href
  {http://dx.doi.org/#2} {doi:#2}\else \href {http://dx.doi.org/#2} {#1}\fi
  \endgroup}
\def\mn@eprint#1#2{\mn@eprint@#1:#2::\@nil}
\def\mn@eprint@arXiv#1{\href {http://arxiv.org/abs/#1} {{\tt arXiv:#1}}}
\def\mn@eprint@dblp#1{\href {http://dblp.uni-trier.de/rec/bibtex/#1.xml}
  {dblp:#1}}
\def\mn@eprint@#1:#2:#3:#4\@nil{\def\@tempa {#1}\def\@tempb {#2}\def\@tempc
  {#3}\ifx \@tempc \@empty \let \@tempc \@tempb \let \@tempb \@tempa \fi \ifx
  \@tempb \@empty \def\@tempb {arXiv}\fi \@ifundefined
  {mn@eprint@\@tempb}{\@tempb:\@tempc}{\expandafter \expandafter \csname
  mn@eprint@\@tempb\endcsname \expandafter{\@tempc}}}

\bibitem[\protect\citeauthoryear{{Ajani}, {Peel}, {Pettorino}, {Starck}, {Li}
  \& {Liu}}{{Ajani} et~al.}{2020}]{Ajani2020}
{Ajani} V.,  {Peel} A.,  {Pettorino} V.,  {Starck} J.-L.,  {Li} Z.,   {Liu} J.,
   2020, \mn@doi [\prd] {10.1103/PhysRevD.102.103531}, \href
  {https://ui.adsabs.harvard.edu/abs/2020PhRvD.102j3531A} {102, 103531}

\bibitem[\protect\citeauthoryear{{Ajani}, {Starck}  \& {Pettorino}}{{Ajani}
  et~al.}{2021}]{Ajani2021}
{Ajani} V.,  {Starck} J.-L.,   {Pettorino} V.,  2021, \mn@doi [\aap]
  {10.1051/0004-6361/202039988}, \href
  {https://ui.adsabs.harvard.edu/abs/2021A&A...645L..11A} {645, L11}

\bibitem[\protect\citeauthoryear{{Alsing}, {Heavens}, {Jaffe}, {Kiessling},
  {Wandelt}  \& {Hoffmann}}{{Alsing} et~al.}{2016}]{Alsing2016}
{Alsing} J.,  {Heavens} A.,  {Jaffe} A.~H.,  {Kiessling} A.,  {Wandelt} B.,
  {Hoffmann} T.,  2016, \mn@doi [\mnras] {10.1093/mnras/stv2501}, \href
  {https://ui.adsabs.harvard.edu/abs/2016MNRAS.455.4452A} {455, 4452}

\bibitem[\protect\citeauthoryear{{Alsing}, {Heavens}  \& {Jaffe}}{{Alsing}
  et~al.}{2017}]{Alsing2017}
{Alsing} J.,  {Heavens} A.,   {Jaffe} A.~H.,  2017, \mn@doi [\mnras]
  {10.1093/mnras/stw3161}, \href
  {https://ui.adsabs.harvard.edu/abs/2017MNRAS.466.3272A} {466, 3272}

\bibitem[\protect\citeauthoryear{{Barreira}, {Krause}  \& {Schmidt}}{{Barreira}
  et~al.}{2018}]{Barreira2018}
{Barreira} A.,  {Krause} E.,   {Schmidt} F.,  2018, \mn@doi [\jcap]
  {10.1088/1475-7516/2018/06/015}, \href
  {https://ui.adsabs.harvard.edu/abs/2018JCAP...06..015B} {2018, 015}

\bibitem[\protect\citeauthoryear{{B{\"o}hm}, {Hilbert}, {Greiner}  \&
  {En{\ss}lin}}{{B{\"o}hm} et~al.}{2017}]{Bohm2017}
{B{\"o}hm} V.,  {Hilbert} S.,  {Greiner} M.,   {En{\ss}lin} T.~A.,  2017,
  \mn@doi [\prd] {10.1103/PhysRevD.96.123510}, \href
  {https://ui.adsabs.harvard.edu/abs/2017PhRvD..96l3510B} {96, 123510}

\bibitem[\protect\citeauthoryear{{Boruah}, {Lavaux}  \& {Hudson}}{{Boruah}
  et~al.}{2021}]{Boruah2021}
{Boruah} S.~S.,  {Lavaux} G.,   {Hudson} M.~J.,  2021, arXiv e-prints, \href
  {https://ui.adsabs.harvard.edu/abs/2021arXiv211115535B} {p. arXiv:2111.15535}

\bibitem[\protect\citeauthoryear{{Boyle}, {Uhlemann}, {Friedrich},
  {Barthelemy}, {Codis}, {Bernardeau}, {Giocoli}  \& {Baldi}}{{Boyle}
  et~al.}{2021}]{Boyle2021}
{Boyle} A.,  {Uhlemann} C.,  {Friedrich} O.,  {Barthelemy} A.,  {Codis} S.,
  {Bernardeau} F.,  {Giocoli} C.,   {Baldi} M.,  2021, \mn@doi [\mnras]
  {10.1093/mnras/stab1381}, \href
  {https://ui.adsabs.harvard.edu/abs/2021MNRAS.505.2886B} {505, 2886}

\bibitem[\protect\citeauthoryear{Bradbury et~al.,}{Bradbury
  et~al.}{2018}]{jax2018github}
Bradbury J.,  et~al., 2018, {JAX}: composable transformations of
  {P}ython+{N}um{P}y programs, \url {http://github.com/google/jax}

\bibitem[\protect\citeauthoryear{{Cheng} \& {M{\'e}nard}}{{Cheng} \&
  {M{\'e}nard}}{2021}]{Cheng2021}
{Cheng} S.,  {M{\'e}nard} B.,  2021, \mn@doi [\mnras] {10.1093/mnras/stab2102},
  \href {https://ui.adsabs.harvard.edu/abs/2021MNRAS.507.1012C} {507, 1012}

\bibitem[\protect\citeauthoryear{{Cheng}, {Ting}, {M{\'e}nard}  \&
  {Bruna}}{{Cheng} et~al.}{2020}]{Cheng2020}
{Cheng} S.,  {Ting} Y.-S.,  {M{\'e}nard} B.,   {Bruna} J.,  2020, \mn@doi
  [\mnras] {10.1093/mnras/staa3165}, \href
  {https://ui.adsabs.harvard.edu/abs/2020MNRAS.499.5902C} {499, 5902}

\bibitem[\protect\citeauthoryear{{Clerkin} et~al.,}{{Clerkin}
  et~al.}{2017}]{Clerkin2017}
{Clerkin} L.,  et~al., 2017, \mn@doi [\mnras] {10.1093/mnras/stw2106}, \href
  {https://ui.adsabs.harvard.edu/abs/2017MNRAS.466.1444C} {466, 1444}

\bibitem[\protect\citeauthoryear{{Coles} \& {Jones}}{{Coles} \&
  {Jones}}{1991}]{Coles1991}
{Coles} P.,  {Jones} B.,  1991, \mn@doi [\mnras] {10.1093/mnras/248.1.1}, \href
  {https://ui.adsabs.harvard.edu/abs/1991MNRAS.248....1C} {248, 1}

\bibitem[\protect\citeauthoryear{{Cragg}, {Duncan}, {Miller}  \&
  {Alonso}}{{Cragg} et~al.}{2022}]{Cragg2022}
{Cragg} C.,  {Duncan} C. A.~J.,  {Miller} L.,   {Alonso} D.,  2022, arXiv
  e-prints, \href {https://ui.adsabs.harvard.edu/abs/2022arXiv220301460C} {p.
  arXiv:2203.01460}

\bibitem[\protect\citeauthoryear{{Dai} \& {Seljak}}{{Dai} \&
  {Seljak}}{2022}]{Dai2022}
{Dai} B.,  {Seljak} U.,  2022, arXiv e-prints, \href
  {https://ui.adsabs.harvard.edu/abs/2022arXiv220205282D} {p. arXiv:2202.05282}

\bibitem[\protect\citeauthoryear{{Elsner} \& {Wandelt}}{{Elsner} \&
  {Wandelt}}{2013}]{Elsner2013}
{Elsner} F.,  {Wandelt} B.~D.,  2013, \mn@doi [\aap]
  {10.1051/0004-6361/201220586}, \href
  {https://ui.adsabs.harvard.edu/abs/2013A&A...549A.111E} {549, A111}

\bibitem[\protect\citeauthoryear{{Euclid Collaboration} et~al.,}{{Euclid
  Collaboration} et~al.}{2019}]{Knabenhans2019}
{Euclid Collaboration} et~al., 2019, \mn@doi [\mnras] {10.1093/mnras/stz197},
  \href {https://ui.adsabs.harvard.edu/abs/2019MNRAS.484.5509E} {484, 5509}

\bibitem[\protect\citeauthoryear{{Fiedorowicz}, {Rozo}, {Boruah}, {Chang}  \&
  {Gatti}}{{Fiedorowicz} et~al.}{2022}]{Fiedorowicz2022}
{Fiedorowicz} P.,  {Rozo} E.,  {Boruah} S.~S.,  {Chang} C.,   {Gatti} M.,
  2022, \mn@doi [\mnras] {10.1093/mnras/stac468}, \href
  {https://ui.adsabs.harvard.edu/abs/2022MNRAS.512...73F} {512, 73}

\bibitem[\protect\citeauthoryear{{Fluri}, {Kacprzak}, {Lucchi}, {Refregier},
  {Amara}, {Hofmann}  \& {Schneider}}{{Fluri} et~al.}{2019}]{Fluri2019}
{Fluri} J.,  {Kacprzak} T.,  {Lucchi} A.,  {Refregier} A.,  {Amara} A.,
  {Hofmann} T.,   {Schneider} A.,  2019, \mn@doi [\prd]
  {10.1103/PhysRevD.100.063514}, \href
  {https://ui.adsabs.harvard.edu/abs/2019PhRvD.100f3514F} {100, 063514}

\bibitem[\protect\citeauthoryear{{Fluri}, {Kacprzak}, {Lucchi}, {Schneider},
  {Refregier}  \& {Hofmann}}{{Fluri} et~al.}{2022}]{Fluri2022}
{Fluri} J.,  {Kacprzak} T.,  {Lucchi} A.,  {Schneider} A.,  {Refregier} A.,
  {Hofmann} T.,  2022, arXiv e-prints, \href
  {https://ui.adsabs.harvard.edu/abs/2022arXiv220107771F} {p. arXiv:2201.07771}

\bibitem[\protect\citeauthoryear{{Friedrich} et~al.,}{{Friedrich}
  et~al.}{2018}]{Friedrich2018}
{Friedrich} O.,  et~al., 2018, \mn@doi [\prd] {10.1103/PhysRevD.98.023508},
  \href {https://ui.adsabs.harvard.edu/abs/2018PhRvD..98b3508F} {98, 023508}

\bibitem[\protect\citeauthoryear{{Friedrich}, {Uhlemann},
  {Villaescusa-Navarro}, {Baldauf}, {Manera}  \& {Nishimichi}}{{Friedrich}
  et~al.}{2020}]{Friedrich2020}
{Friedrich} O.,  {Uhlemann} C.,  {Villaescusa-Navarro} F.,  {Baldauf} T.,
  {Manera} M.,   {Nishimichi} T.,  2020, \mn@doi [\mnras]
  {10.1093/mnras/staa2160}, \href
  {https://ui.adsabs.harvard.edu/abs/2020MNRAS.498..464F} {498, 464}

\bibitem[\protect\citeauthoryear{{Gatti} et~al.,}{{Gatti}
  et~al.}{2020}]{Gatti2020}
{Gatti} M.,  et~al., 2020, \mn@doi [\mnras] {10.1093/mnras/staa2680}, \href
  {https://ui.adsabs.harvard.edu/abs/2020MNRAS.498.4060G} {498, 4060}

\bibitem[\protect\citeauthoryear{{Gatti} et~al.,}{{Gatti}
  et~al.}{2021}]{Gatti2021}
{Gatti} M.,  et~al., 2021, arXiv e-prints, \href
  {https://ui.adsabs.harvard.edu/abs/2021arXiv211010141G} {p. arXiv:2110.10141}

\bibitem[\protect\citeauthoryear{Gibbons \& Chakraborti}{Gibbons \&
  Chakraborti}{2011}]{Gibbons2011}
Gibbons J.~D.,  Chakraborti S.,  2011, Nonparametric Statistical Inference.
Springer Berlin Heidelberg, Berlin, Heidelberg, pp 977--979,
  \mn@doi{10.1007/978-3-642-04898-2_420}, \url
  {https://doi.org/10.1007/978-3-642-04898-2_420}

\bibitem[\protect\citeauthoryear{{Halder}, {Friedrich}, {Seitz}  \&
  {Varga}}{{Halder} et~al.}{2021}]{Halder2021}
{Halder} A.,  {Friedrich} O.,  {Seitz} S.,   {Varga} T.~N.,  2021, \mn@doi
  [\mnras] {10.1093/mnras/stab1801}, \href
  {https://ui.adsabs.harvard.edu/abs/2021MNRAS.506.2780H} {506, 2780}

\bibitem[\protect\citeauthoryear{{Hall} \& {Mead}}{{Hall} \&
  {Mead}}{2018}]{Hall2018}
{Hall} A.,  {Mead} A.,  2018, \mn@doi [\mnras] {10.1093/mnras/stx2575}, \href
  {https://ui.adsabs.harvard.edu/abs/2018MNRAS.473.3190H} {473, 3190}

\bibitem[\protect\citeauthoryear{{Hamana} et~al.,}{{Hamana}
  et~al.}{2020}]{Hamana2020}
{Hamana} T.,  et~al., 2020, \mn@doi [\pasj] {10.1093/pasj/psz138}, \href
  {https://ui.adsabs.harvard.edu/abs/2020PASJ...72...16H} {72, 16}

\bibitem[\protect\citeauthoryear{{Harnois-D{\'e}raps}, {Martinet}, {Castro},
  {Dolag}, {Giblin}, {Heymans}, {Hildebrandt}  \& {Xia}}{{Harnois-D{\'e}raps}
  et~al.}{2021}]{HarnoisDeraps2021}
{Harnois-D{\'e}raps} J.,  {Martinet} N.,  {Castro} T.,  {Dolag} K.,  {Giblin}
  B.,  {Heymans} C.,  {Hildebrandt} H.,   {Xia} Q.,  2021, \mn@doi [\mnras]
  {10.1093/mnras/stab1623}, \href
  {https://ui.adsabs.harvard.edu/abs/2021MNRAS.506.1623H} {506, 1623}

\bibitem[\protect\citeauthoryear{{Heymans} et~al.,}{{Heymans}
  et~al.}{2021}]{Heymans2021}
{Heymans} C.,  et~al., 2021, \mn@doi [\aap] {10.1051/0004-6361/202039063},
  \href {https://ui.adsabs.harvard.edu/abs/2021A&A...646A.140H} {646, A140}

\bibitem[\protect\citeauthoryear{{Hikage} et~al.,}{{Hikage}
  et~al.}{2019}]{Hikage2019}
{Hikage} C.,  et~al., 2019, \mn@doi [\pasj] {10.1093/pasj/psz010}, \href
  {https://ui.adsabs.harvard.edu/abs/2019PASJ...71...43H} {71, 43}

\bibitem[\protect\citeauthoryear{{Howlett}, {Lewis}, {Hall}  \&
  {Challinor}}{{Howlett} et~al.}{2012}]{Howlett2012}
{Howlett} C.,  {Lewis} A.,  {Hall} A.,   {Challinor} A.,  2012, \mn@doi [\jcap]
  {10.1088/1475-7516/2012/04/027}, \href
  {https://ui.adsabs.harvard.edu/abs/2012JCAP...04..027H} {2012, 027}

\bibitem[\protect\citeauthoryear{{Jasche} \& {Kitaura}}{{Jasche} \&
  {Kitaura}}{2010}]{Jasche2010}
{Jasche} J.,  {Kitaura} F.~S.,  2010, \mn@doi [\mnras]
  {10.1111/j.1365-2966.2010.16897.x}, \href
  {https://ui.adsabs.harvard.edu/abs/2010MNRAS.407...29J} {407, 29}

\bibitem[\protect\citeauthoryear{{Jasche} \& {Lavaux}}{{Jasche} \&
  {Lavaux}}{2019}]{Jasche2019}
{Jasche} J.,  {Lavaux} G.,  2019, \mn@doi [\aap] {10.1051/0004-6361/201833710},
  \href {https://ui.adsabs.harvard.edu/abs/2019A&A...625A..64J} {625, A64}

\bibitem[\protect\citeauthoryear{{Jasche} \& {Wandelt}}{{Jasche} \&
  {Wandelt}}{2013}]{Jasche2013}
{Jasche} J.,  {Wandelt} B.~D.,  2013, \mn@doi [\mnras] {10.1093/mnras/stt449},
  \href {https://ui.adsabs.harvard.edu/abs/2013MNRAS.432..894J} {432, 894}

\bibitem[\protect\citeauthoryear{{Jeffrey}, {Alsing}  \& {Lanusse}}{{Jeffrey}
  et~al.}{2021}]{Jeffrey2021}
{Jeffrey} N.,  {Alsing} J.,   {Lanusse} F.,  2021, \mn@doi [\mnras]
  {10.1093/mnras/staa3594}, \href
  {https://ui.adsabs.harvard.edu/abs/2021MNRAS.501..954J} {501, 954}

\bibitem[\protect\citeauthoryear{{Kaiser} \& {Squires}}{{Kaiser} \&
  {Squires}}{1993}]{Kaiser1993}
{Kaiser} N.,  {Squires} G.,  1993, \mn@doi [\apj] {10.1086/172297}, \href
  {https://ui.adsabs.harvard.edu/abs/1993ApJ...404..441K} {404, 441}

\bibitem[\protect\citeauthoryear{{Klypin}, {Prada}, {Betancort-Rijo}  \&
  {Albareti}}{{Klypin} et~al.}{2018}]{Klypin2018}
{Klypin} A.,  {Prada} F.,  {Betancort-Rijo} J.,   {Albareti} F.~D.,  2018,
  \mn@doi [\mnras] {10.1093/mnras/sty2613}, \href
  {https://ui.adsabs.harvard.edu/abs/2018MNRAS.481.4588K} {481, 4588}

\bibitem[\protect\citeauthoryear{{Kokron}, {DeRose}, {Chen}, {White}  \&
  {Wechsler}}{{Kokron} et~al.}{2021}]{Kokron2021}
{Kokron} N.,  {DeRose} J.,  {Chen} S.-F.,  {White} M.,   {Wechsler} R.~H.,
  2021, \mn@doi [\mnras] {10.1093/mnras/stab1358}, \href
  {https://ui.adsabs.harvard.edu/abs/2021MNRAS.505.1422K} {505, 1422}

\bibitem[\protect\citeauthoryear{{Kratochvil}, {Lim}, {Wang}, {Haiman}, {May}
  \& {Huffenberger}}{{Kratochvil} et~al.}{2012}]{Kratochvil2012}
{Kratochvil} J.~M.,  {Lim} E.~A.,  {Wang} S.,  {Haiman} Z.,  {May} M.,
  {Huffenberger} K.,  2012, \mn@doi [\prd] {10.1103/PhysRevD.85.103513}, \href
  {https://ui.adsabs.harvard.edu/abs/2012PhRvD..85j3513K} {85, 103513}

\bibitem[\protect\citeauthoryear{{Lanusse}, {Starck}, {Leonard}  \&
  {Pires}}{{Lanusse} et~al.}{2016}]{Lanusse2016}
{Lanusse} F.,  {Starck} J.~L.,  {Leonard} A.,   {Pires} S.,  2016, \mn@doi
  [\aap] {10.1051/0004-6361/201628278}, \href
  {https://ui.adsabs.harvard.edu/abs/2016A&A...591A...2L} {591, A2}

\bibitem[\protect\citeauthoryear{{Leclercq} \& {Heavens}}{{Leclercq} \&
  {Heavens}}{2021}]{Leclercq2021}
{Leclercq} F.,  {Heavens} A.,  2021, \mn@doi [\mnras] {10.1093/mnrasl/slab081},
  \href {https://ui.adsabs.harvard.edu/abs/2021MNRAS.506L..85L} {506, L85}

\bibitem[\protect\citeauthoryear{{Lewis}, {Challinor}  \& {Lasenby}}{{Lewis}
  et~al.}{2000}]{Lewis2000}
{Lewis} A.,  {Challinor} A.,   {Lasenby} A.,  2000, \mn@doi [\apj]
  {10.1086/309179}, \href
  {https://ui.adsabs.harvard.edu/abs/2000ApJ...538..473L} {538, 473}

\bibitem[\protect\citeauthoryear{{Liu} \& {Madhavacheril}}{{Liu} \&
  {Madhavacheril}}{2019}]{Liu2019}
{Liu} J.,  {Madhavacheril} M.~S.,  2019, \mn@doi [\prd]
  {10.1103/PhysRevD.99.083508}, \href
  {https://ui.adsabs.harvard.edu/abs/2019PhRvD..99h3508L} {99, 083508}

\bibitem[\protect\citeauthoryear{{Liu}, {Petri}, {Haiman}, {Hui}, {Kratochvil}
  \& {May}}{{Liu} et~al.}{2015}]{Liu2015}
{Liu} J.,  {Petri} A.,  {Haiman} Z.,  {Hui} L.,  {Kratochvil} J.~M.,   {May}
  M.,  2015, \mn@doi [\prd] {10.1103/PhysRevD.91.063507}, \href
  {https://ui.adsabs.harvard.edu/abs/2015PhRvD..91f3507L} {91, 063507}

\bibitem[\protect\citeauthoryear{{Marques}, {Liu}, {Zorrilla Matilla},
  {Haiman}, {Bernui}  \& {Novaes}}{{Marques} et~al.}{2019}]{Marques2019}
{Marques} G.~A.,  {Liu} J.,  {Zorrilla Matilla} J.~M.,  {Haiman} Z.,  {Bernui}
  A.,   {Novaes} C.~P.,  2019, \mn@doi [\jcap] {10.1088/1475-7516/2019/06/019},
  \href {https://ui.adsabs.harvard.edu/abs/2019JCAP...06..019M} {2019, 019}

\bibitem[\protect\citeauthoryear{{Martinet}, {Harnois-D{\'e}raps}, {Jullo}  \&
  {Schneider}}{{Martinet} et~al.}{2021}]{Martinet2021}
{Martinet} N.,  {Harnois-D{\'e}raps} J.,  {Jullo} E.,   {Schneider} P.,  2021,
  \mn@doi [\aap] {10.1051/0004-6361/202039679}, \href
  {https://ui.adsabs.harvard.edu/abs/2021A&A...646A..62M} {646, A62}

\bibitem[\protect\citeauthoryear{{Matilla}, {Sharma}, {Hsu}  \&
  {Haiman}}{{Matilla} et~al.}{2020}]{Matilla2020}
{Matilla} J. M.~Z.,  {Sharma} M.,  {Hsu} D.,   {Haiman} Z.,  2020, \mn@doi
  [\prd] {10.1103/PhysRevD.102.123506}, \href
  {https://ui.adsabs.harvard.edu/abs/2020PhRvD.102l3506M} {102, 123506}

\bibitem[\protect\citeauthoryear{{Millea}, {Anderes}  \& {Wandelt}}{{Millea}
  et~al.}{2020}]{Millea2020b}
{Millea} M.,  {Anderes} E.,   {Wandelt} B.~D.,  2020, \mn@doi [\prd]
  {10.1103/PhysRevD.102.123542}, \href
  {https://ui.adsabs.harvard.edu/abs/2020PhRvD.102l3542M} {102, 123542}

\bibitem[\protect\citeauthoryear{{Mustafa}, {Bard}, {Bhimji}, {Luki{\'c}},
  {Al-Rfou}  \& {Kratochvil}}{{Mustafa} et~al.}{2019}]{Mustafa2019}
{Mustafa} M.,  {Bard} D.,  {Bhimji} W.,  {Luki{\'c}} Z.,  {Al-Rfou} R.,
  {Kratochvil} J.~M.,  2019, \mn@doi [Computational Astrophysics and Cosmology]
  {10.1186/s40668-019-0029-9}, \href
  {https://ui.adsabs.harvard.edu/abs/2019ComAC...6....1M} {6, 1}

\bibitem[\protect\citeauthoryear{{Neal}}{{Neal}}{2000}]{Neal2000}
{Neal} R.~M.,  2000, arXiv e-prints, \href
  {https://ui.adsabs.harvard.edu/abs/2000physics...9028N} {p. physics/0009028}

\bibitem[\protect\citeauthoryear{{Neal}}{{Neal}}{2012}]{Neal2012}
{Neal} R.~M.,  2012, arXiv e-prints, \href
  {https://ui.adsabs.harvard.edu/abs/2012arXiv1206.1901N} {p. arXiv:1206.1901}

\bibitem[\protect\citeauthoryear{{Nguyen}, {Jasche}, {Lavaux}  \&
  {Schmidt}}{{Nguyen} et~al.}{2020}]{Nguyen2020}
{Nguyen} N.-M.,  {Jasche} J.,  {Lavaux} G.,   {Schmidt} F.,  2020, \mn@doi
  [\jcap] {10.1088/1475-7516/2020/12/011}, \href
  {https://ui.adsabs.harvard.edu/abs/2020JCAP...12..011N} {2020, 011}

\bibitem[\protect\citeauthoryear{{Perraudin}, {Marcon}, {Lucchi}  \&
  {Kacprzak}}{{Perraudin} et~al.}{2020}]{Perraudin2020}
{Perraudin} N.,  {Marcon} S.,  {Lucchi} A.,   {Kacprzak} T.,  2020, arXiv
  e-prints, \href {https://ui.adsabs.harvard.edu/abs/2020arXiv200408139P} {p.
  arXiv:2004.08139}

\bibitem[\protect\citeauthoryear{{Petri}, {Haiman}, {Hui}, {May}  \&
  {Kratochvil}}{{Petri} et~al.}{2013}]{Petri2013}
{Petri} A.,  {Haiman} Z.,  {Hui} L.,  {May} M.,   {Kratochvil} J.~M.,  2013,
  \mn@doi [\prd] {10.1103/PhysRevD.88.123002}, \href
  {https://ui.adsabs.harvard.edu/abs/2013PhRvD..88l3002P} {88, 123002}

\bibitem[\protect\citeauthoryear{{Porqueres}, {Heavens}, {Mortlock}  \&
  {Lavaux}}{{Porqueres} et~al.}{2021}]{Porqueres2021}
{Porqueres} N.,  {Heavens} A.,  {Mortlock} D.,   {Lavaux} G.,  2021, \mn@doi
  [\mnras] {10.1093/mnras/stab204}, \href
  {https://ui.adsabs.harvard.edu/abs/2021MNRAS.502.3035P} {502, 3035}

\bibitem[\protect\citeauthoryear{{Porqueres}, {Heavens}, {Mortlock}  \&
  {Lavaux}}{{Porqueres} et~al.}{2022}]{Porqueres2022}
{Porqueres} N.,  {Heavens} A.,  {Mortlock} D.,   {Lavaux} G.,  2022, \mn@doi
  [\mnras] {10.1093/mnras/stab3234}, \href
  {https://ui.adsabs.harvard.edu/abs/2022MNRAS.509.3194P} {509, 3194}

\bibitem[\protect\citeauthoryear{{Remy}, {Lanusse}, {Ramzi}, {Liu}, {Jeffrey}
  \& {Starck}}{{Remy} et~al.}{2020}]{Remy2020}
{Remy} B.,  {Lanusse} F.,  {Ramzi} Z.,  {Liu} J.,  {Jeffrey} N.,   {Starck}
  J.-L.,  2020, arXiv e-prints, \href
  {https://ui.adsabs.harvard.edu/abs/2020arXiv201108271R} {p. arXiv:2011.08271}

\bibitem[\protect\citeauthoryear{{Remy}, {Lanusse}, {Jeffrey}, {Liu}, {Starck},
  {Osato}  \& {Schrabback}}{{Remy} et~al.}{2022}]{Remy2022}
{Remy} B.,  {Lanusse} F.,  {Jeffrey} N.,  {Liu} J.,  {Starck} J.-L.,  {Osato}
  K.,   {Schrabback} T.,  2022, arXiv e-prints, \href
  {https://ui.adsabs.harvard.edu/abs/2022arXiv220105561R} {p. arXiv:2201.05561}

\bibitem[\protect\citeauthoryear{{Ribli}, {Pataki}, {Zorrilla Matilla}, {Hsu},
  {Haiman}  \& {Csabai}}{{Ribli} et~al.}{2019}]{Ribli2019}
{Ribli} D.,  {Pataki} B.~{\'A}.,  {Zorrilla Matilla} J.~M.,  {Hsu} D.,
  {Haiman} Z.,   {Csabai} I.,  2019, \mn@doi [\mnras] {10.1093/mnras/stz2610},
  \href {https://ui.adsabs.harvard.edu/abs/2019MNRAS.490.1843R} {490, 1843}

\bibitem[\protect\citeauthoryear{{Sellentin} \& {Heavens}}{{Sellentin} \&
  {Heavens}}{2016}]{Sellentin2016}
{Sellentin} E.,  {Heavens} A.~F.,  2016, \mn@doi [\mnras]
  {10.1093/mnrasl/slv190}, \href
  {https://ui.adsabs.harvard.edu/abs/2016MNRAS.456L.132S} {456, L132}

\bibitem[\protect\citeauthoryear{{Semboloni}, {Schrabback}, {van Waerbeke},
  {Vafaei}, {Hartlap}  \& {Hilbert}}{{Semboloni} et~al.}{2011}]{Semboloni2011}
{Semboloni} E.,  {Schrabback} T.,  {van Waerbeke} L.,  {Vafaei} S.,  {Hartlap}
  J.,   {Hilbert} S.,  2011, \mn@doi [\mnras]
  {10.1111/j.1365-2966.2010.17430.x}, \href
  {https://ui.adsabs.harvard.edu/abs/2011MNRAS.410..143S} {410, 143}

\bibitem[\protect\citeauthoryear{{Smail}, {Hogg}, {Yan}  \& {Cohen}}{{Smail}
  et~al.}{1995}]{Smail1995}
{Smail} I.,  {Hogg} D.~W.,  {Yan} L.,   {Cohen} J.~G.,  1995, \mn@doi [\apjl]
  {10.1086/309647}, \href
  {https://ui.adsabs.harvard.edu/abs/1995ApJ...449L.105S} {449, L105}

\bibitem[\protect\citeauthoryear{{Starck}, {Themelis}, {Jeffrey}, {Peel}  \&
  {Lanusse}}{{Starck} et~al.}{2021}]{Starck2021}
{Starck} J.~L.,  {Themelis} K.~E.,  {Jeffrey} N.,  {Peel} A.,   {Lanusse} F.,
  2021, \mn@doi [\aap] {10.1051/0004-6361/202039451}, \href
  {https://ui.adsabs.harvard.edu/abs/2021A&A...649A..99S} {649, A99}

\bibitem[\protect\citeauthoryear{{Takada} \& {Jain}}{{Takada} \&
  {Jain}}{2004}]{Takada2004}
{Takada} M.,  {Jain} B.,  2004, \mn@doi [\mnras]
  {10.1111/j.1365-2966.2004.07410.x}, \href
  {https://ui.adsabs.harvard.edu/abs/2004MNRAS.348..897T} {348, 897}

\bibitem[\protect\citeauthoryear{{Taruya}, {Takada}, {Hamana}, {Kayo}  \&
  {Futamase}}{{Taruya} et~al.}{2002}]{Taruya2002}
{Taruya} A.,  {Takada} M.,  {Hamana} T.,  {Kayo} I.,   {Futamase} T.,  2002,
  \mn@doi [\apj] {10.1086/340048}, \href
  {https://ui.adsabs.harvard.edu/abs/2002ApJ...571..638T} {571, 638}

\bibitem[\protect\citeauthoryear{{Taylor}, {Ashdown}  \& {Hobson}}{{Taylor}
  et~al.}{2008}]{Taylor2008}
{Taylor} J.~F.,  {Ashdown} M.~A.~J.,   {Hobson} M.~P.,  2008, \mn@doi [\mnras]
  {10.1111/j.1365-2966.2008.13630.x}, \href
  {https://ui.adsabs.harvard.edu/abs/2008MNRAS.389.1284T} {389, 1284}

\bibitem[\protect\citeauthoryear{{The LSST Dark Energy Science Collaboration}
  et~al.,}{{The LSST Dark Energy Science Collaboration}
  et~al.}{2018}]{LSST_SRD}
{The LSST Dark Energy Science Collaboration} et~al., 2018, arXiv e-prints,
  \href {https://ui.adsabs.harvard.edu/abs/2018arXiv180901669T} {p.
  arXiv:1809.01669}

\bibitem[\protect\citeauthoryear{{Thiele}, {Hill}  \& {Smith}}{{Thiele}
  et~al.}{2020}]{Thiele2020}
{Thiele} L.,  {Hill} J.~C.,   {Smith} K.~M.,  2020, \mn@doi [\prd]
  {10.1103/PhysRevD.102.123545}, \href
  {https://ui.adsabs.harvard.edu/abs/2020PhRvD.102l3545T} {102, 123545}

\bibitem[\protect\citeauthoryear{{Troxel} et~al.,}{{Troxel}
  et~al.}{2018}]{Troxel2018}
{Troxel} M.~A.,  et~al., 2018, \mn@doi [\prd] {10.1103/PhysRevD.98.043528},
  \href {https://ui.adsabs.harvard.edu/abs/2018PhRvD..98d3528T} {98, 043528}

\bibitem[\protect\citeauthoryear{{Tsaprazi}, {Nguyen}, {Jasche}, {Schmidt}  \&
  {Lavaux}}{{Tsaprazi} et~al.}{2021}]{Tsaprazi2021}
{Tsaprazi} E.,  {Nguyen} N.-M.,  {Jasche} J.,  {Schmidt} F.,   {Lavaux} G.,
  2021, arXiv e-prints, \href
  {https://ui.adsabs.harvard.edu/abs/2021arXiv211204484T} {p. arXiv:2112.04484}

\bibitem[\protect\citeauthoryear{{Xavier}, {Abdalla}  \& {Joachimi}}{{Xavier}
  et~al.}{2016}]{Xavier2016}
{Xavier} H.~S.,  {Abdalla} F.~B.,   {Joachimi} B.,  2016, \mn@doi [\mnras]
  {10.1093/mnras/stw874}, \href
  {https://ui.adsabs.harvard.edu/abs/2016MNRAS.459.3693X} {459, 3693}

\bibitem[\protect\citeauthoryear{{Z{\"u}rcher} et~al.,}{{Z{\"u}rcher}
  et~al.}{2022}]{Zurcher2022}
{Z{\"u}rcher} D.,  et~al., 2022, \mn@doi [\mnras] {10.1093/mnras/stac078},
  \href {https://ui.adsabs.harvard.edu/abs/2022MNRAS.511.2075Z} {511, 2075}

\makeatother
\end{thebibliography}

%%%%%%%%%%%%%%%%%%%%%%%%%%%%%%%%%%%%%%%%%%%%%%%%%%

%%%%%%%%%%%%%%%%% APPENDICES %%%%%%%%%%%%%%%%%%%%%
\appendix
% \input{sections/appendix/emulator}
% \input{sections/appendix/analytic_posterior}
% \input{sections/appendix/LN_shift_calculation}
% \input{sections/appendix/sampling_efficiency}
% Don't change these lines
\bsp	% typesetting comment
\label{lastpage}
\end{document}